\documentclass[pdflatex,sn-mathphys-num, twocolumn]{sn-jnl}% Math and Physical Sciences Numbered Reference Style
\usepackage{graphicx}%
\usepackage{multirow}%
\usepackage{amsmath,amssymb,amsfonts}%
\usepackage{amsthm}%
\usepackage{mathrsfs}%
\usepackage[title]{appendix}%
\usepackage{xcolor}%
\usepackage{textcomp}%
\usepackage{manyfoot}%
\usepackage{booktabs}%
\usepackage{algorithm}%
\usepackage{algorithmicx}%
\usepackage{algpseudocode}%
\usepackage{listings}%
\usepackage{tikz}
\usepackage{cancel}

\usepackage{array}
\usepackage{caption}
\usepackage{booktabs}
\usepackage{graphicx}
\usepackage{geometry}
\usepackage{graphicx}
\usepackage{subcaption}
\usepackage[switch]{lineno}
\theoremstyle{thmstyleone}%
\theoremstyle{thmstyletwo}%

\theoremstyle{thmstylethree}%

\begin{document}

%\title{Reconstructing the dark Universe in three dimensions with JWST}

%\title{Tracing the growth of the cosmic web with JWST}
%\title{A high-resolution three-dimensional generative reconstruction of dark matter}
%\title{Unveiling the dark cosmic web in three dimensions}
%\title{Three-dimensional weak lensing mapping of the dark cosmic web with JWST}
%\title{Generative three-dimensional mapping of the dark cosmic web with JWST}

%\title{The COSMOS-Web 3D dark matter map: A generative reconstruction with JWST}
\title{3D dark matter map of the JWST COSMOS-Web field using diffusion-based generative priors}

\author[1,2]{\fnm{Diana} \sur{Scognamiglio}}
\equalcont{These authors contributed equally to this work.}

\author[3]{\fnm{Brandon} \sur{Zhao}}
\equalcont{These authors contributed equally to this work.}

\author[3, 4]{\fnm{Katherine L.} \sur{Bouman}}
\author[2, 4]{\fnm{Olivier} \sur{Dor\'e }}
\author[5, 6, 7]{\fnm{Caitlin M.} \sur{Casey}}
\author[6, 8]{\fnm{Maximilien} \sur{Franco}}
\author[9, 10]{\fnm{Ghassem} \sur{Gozaliasl}}
\author[11]{\fnm{David} \sur{Harvey}}
\author[12]{\fnm{Olivier} \sur{Ilbert}}
\author[2]{\fnm{Jeffrey B.} \sur{Jewell}}
\author[13]{\fnm{Jeyhan S.} \sur{Kartaltepe}}
\author[14]{\fnm{Anton M.} \sur{Koekemoer}} 
\author[15]{\fnm{Gavin} \sur{Leroy}}
\author[15]{\fnm{Richard} \sur{Massey}}
\author[2]{\fnm{Jason} \sur{Rhodes}}
\author[16]{\fnm{Brant E. } \sur{Robertson}}
\author[7]{\fnm{Marko} \sur{Shuntov}}
\author[7]{\fnm{Sune} \sur{Toft}}
\author[17, 18]{\fnm{Greta} \sur{Toni}}
\author[1]{\fnm{Michael A.} \sur{Troxel}}

%\author[]{\fnm{\& Co}}
% \author[1,2]{\fnm{Diana} \sur{Scognamiglio}}
% %\equalcont{These authors contributed equally to this work.}
% \author[3]{\fnm{Brandon} \sur{Zhao}}
% %\equalcont{These authors contributed equally to this work.}
% \author[3]{\fnm{xx} \sur{xx}}
% \author[2]{\fnm{xx} \sur{xx}}

\affil[1]{Duke University, Durham, NC 27708, USA}
\affil[2]{Jet Propulsion Laboratory, California Institute of Technology, 4800 Oak Grove Drive, Pasadena, CA 91109, USA}
\affil[3]{Department of Computing and Mathematical Sciences, California Institute of Technology}
\affil[4]{Cahill Center for Astronomy and Astrophysics, California Institute of Technology}

\affil[5]{Department of Physics, University of California Santa Barbara, Santa Barbara, CA 93106, USA}
\affil[6]{The University of Texas at Austin, 2515 Speedway Blvd Stop C1400, Austin, TX 78712, USA}
\affil[7]{Cosmic Dawn Center (DAWN), Denmark}
\affil[8]{Université Paris-Saclay, Université Paris Cité, CEA, CNRS, AIM, 91191 Gif-sur-Yvette, France}
\affil[9]{Department of Computer Science, Aalto University, PO Box 15400, Espoo, FI-00 076, Finland}
\affil[10]{Department of Physics, Faculty of Science, University of Helsinki, 00014 Helsinki, Finland}
\affil[11]{Laboratoire d’Astrophysique, École Polytechnique Fédérale de Lausanne (EPFL), Observatoire de Sauverny, CH-1290 Versoix, Switzerland}
\affil[12]{Aix Marseille Univ., CNRS, CNES, LAM, Marseille, France}
\affil[13]{Laboratory for Multiwavelength Astrophysics, School of Physics and Astronomy, Rochester Institute of Technology, 84 Lomb Memorial Drive, Rochester, NY 14623, USA}
\affil[14]{Space Telescope Science Institute, 3700 San Martin Drive, Baltimore, MD 21218, USA}
\affil[15]{Institute for Computational Cosmology, Department of Physics, Durham University, South Road, Durham DH1 3LE, United Kingdom}
\affil[16]{Department of Astronomy and Astrophysics, University of California, Santa Cruz, 1156 High Street, Santa Cruz, CA 95064, USA}
\affil[17]{Dipartimento di Fisica e Astronomia “A. Righi”, Alma Mater Studiorum Università di Bologna, via Gobetti 93/2, 40129 Bologna, Italy}
\affil[18]{INAF – Osservatorio di Astrofisica e Scienza dello Spazio di Bologna, via Gobetti 93/3, 40129 Bologna, Italy}

%%%NEW ABSTRACT OF 183 WORDS IN ORDER TO BE COMPLIANT WITH THE STYLE POLICY%%%
%Before resubmission:
\abstract{Mapping the three-dimensional distribution of dark matter is key to tracing the growth of cosmic structure. Two-dimensional weak lensing mass maps recover its projected distribution, but they blend structures at different distances along the line of sight. We reconstruct the three-dimensional matter distribution in the 0.54 deg$^{2}$ COSMOS-Web field using deep \textit{James Webb} Space Telescope imaging, with an effective source density of 121 galaxies arcmin$^{-2}$. We combine a physical weak lensing forward model with a diffusion-based Bayesian generative prior learned from cosmological simulations, yielding a posterior distribution of the matter field across 20 lens planes to $z\simeq2$, doubling the redshift reach of previous \textit{Hubble} Space Telescope-based 3D reconstructions and approaching cosmic noon, an epoch of rapid galaxy growth. The reconstruction separates structures that overlap in projection and recovers non-Gaussian matter structure across a range of spatial scales, including over-densities associated with independently identified galaxy clusters that were not detected in earlier weak lensing maps. Posterior sampling quantifies the range of three-dimensional matter distributions allowed by the data, enabling a probabilistic visualization of the cosmic web while propagating uncertainties in its inferred structure. Null and simulation-based tests show that the recovered structures are supported by the lensing signal and are consistent with expectations from the fiducial $\Lambda$CDM cosmology. These results extend three-dimensional weak lensing mapping into the high-redshift, non-Gaussian regime and establish a route towards probabilistic mapping of the evolving cosmic web with current wide-field lensing surveys.}

\maketitle
%\linenumbers
\section*{Main}\label{main}
Cosmic structures form a three-dimensional (3D) web of dark-matter haloes, filaments, and voids that grows under gravity from primordial density fluctuations \cite{Springel_2005, Huterer_2023}. Weak gravitational lensing (WL) provides a direct probe of the matter distribution through the coherent distortion of background galaxies \cite{BertSchn01,Kilbinger_2015}. However, two-dimensional (2D) WL  mass maps blend structures at different distances along the line of sight. Recovering where matter lies in three dimensions would provide a direct view of the cosmic web across cosmic time, connecting galaxies, groups and clusters to their dark-matter environments.  On non-linear scales, this distribution encodes the assembly of dark-matter
haloes and their connection to the surrounding cosmic web, but recovering this information requires solving a severely ill-posed inverse problem.\\

%%%%NEW 3D VISUALIZATION POSTERIOR MEAN, BEST SAMPLE, RANDOM SAMPLE%%%%%
\begin{figure*}[t]
\centering
% ============================================================
% Row a: Posterior mean
% ============================================================
\noindent
\begin{minipage}[t]{0.025\textwidth}
\vspace{0pt}
\textsf{\textbf{\large a}}
\end{minipage}%
\hfill
\begin{minipage}[t]{0.965\textwidth}
\vspace{0pt}
\centering
\includegraphics[
    width=0.32\linewidth,
    trim=30 0 30 0,
    clip
]{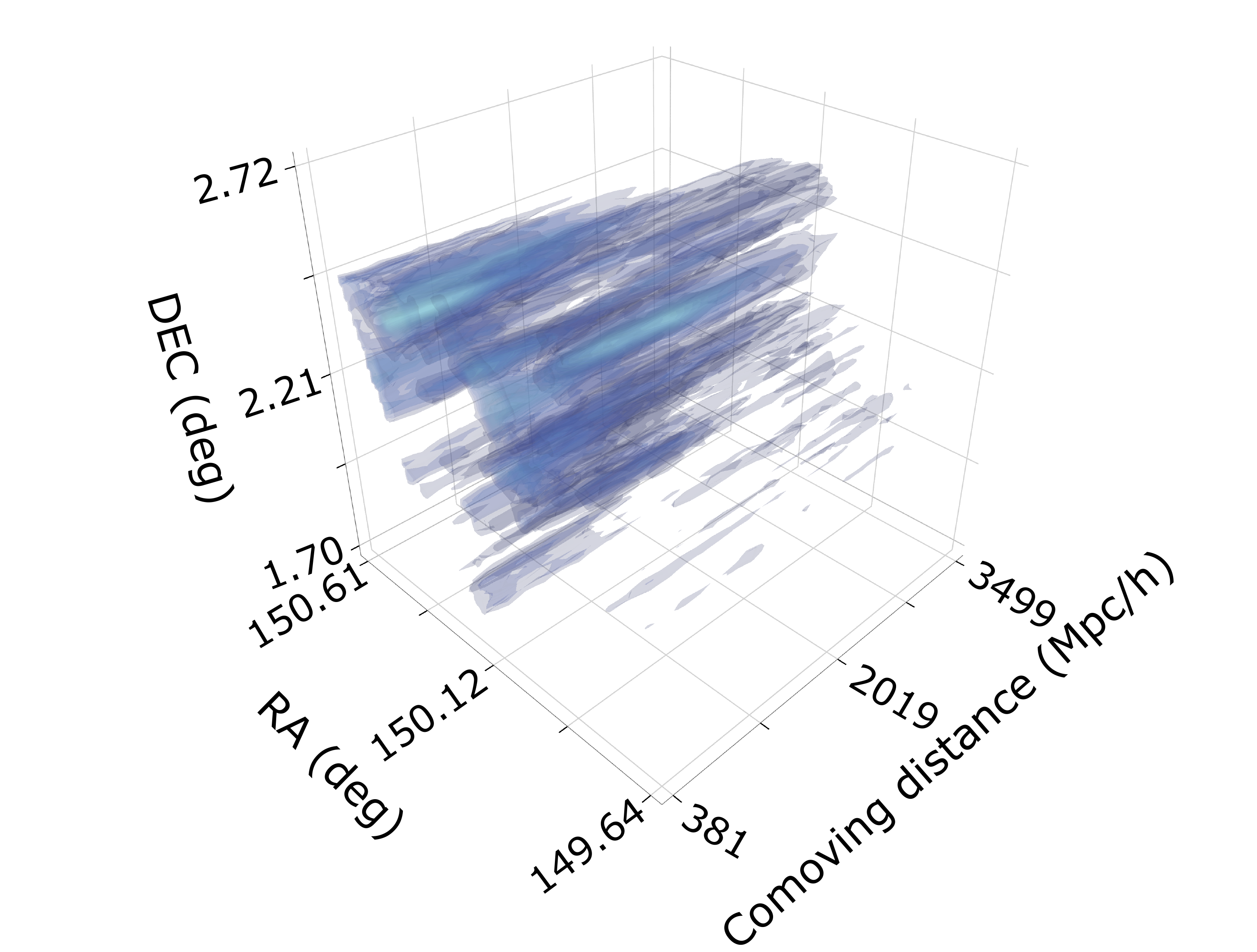}%
\hspace{0.003\linewidth}%
\includegraphics[
    width=0.32\linewidth,
    trim=30 0 30 0,
    clip
]{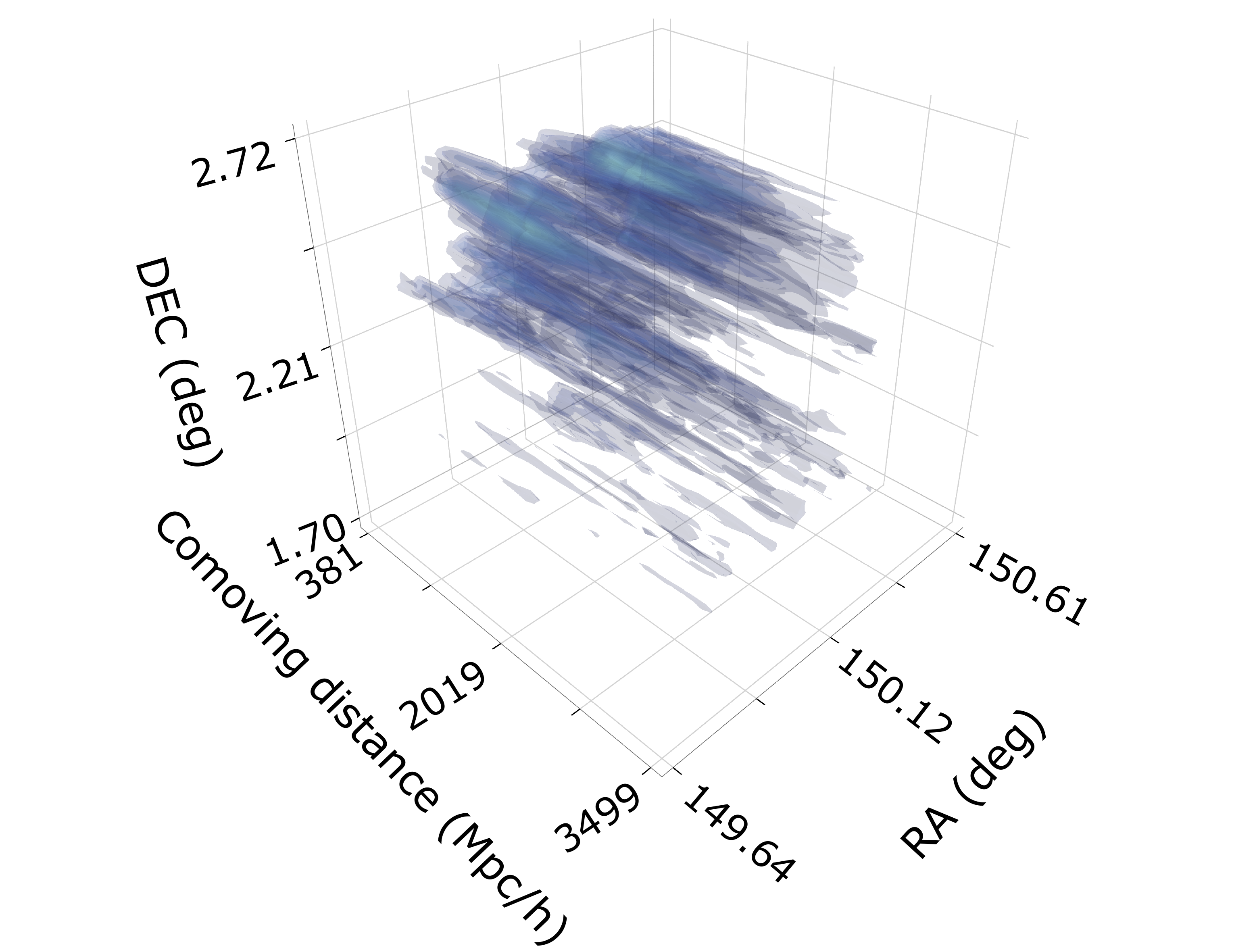}%
\hspace{0.003\linewidth}%
\includegraphics[
    width=0.32\linewidth,
    trim=30 0 30 0,
    clip
]{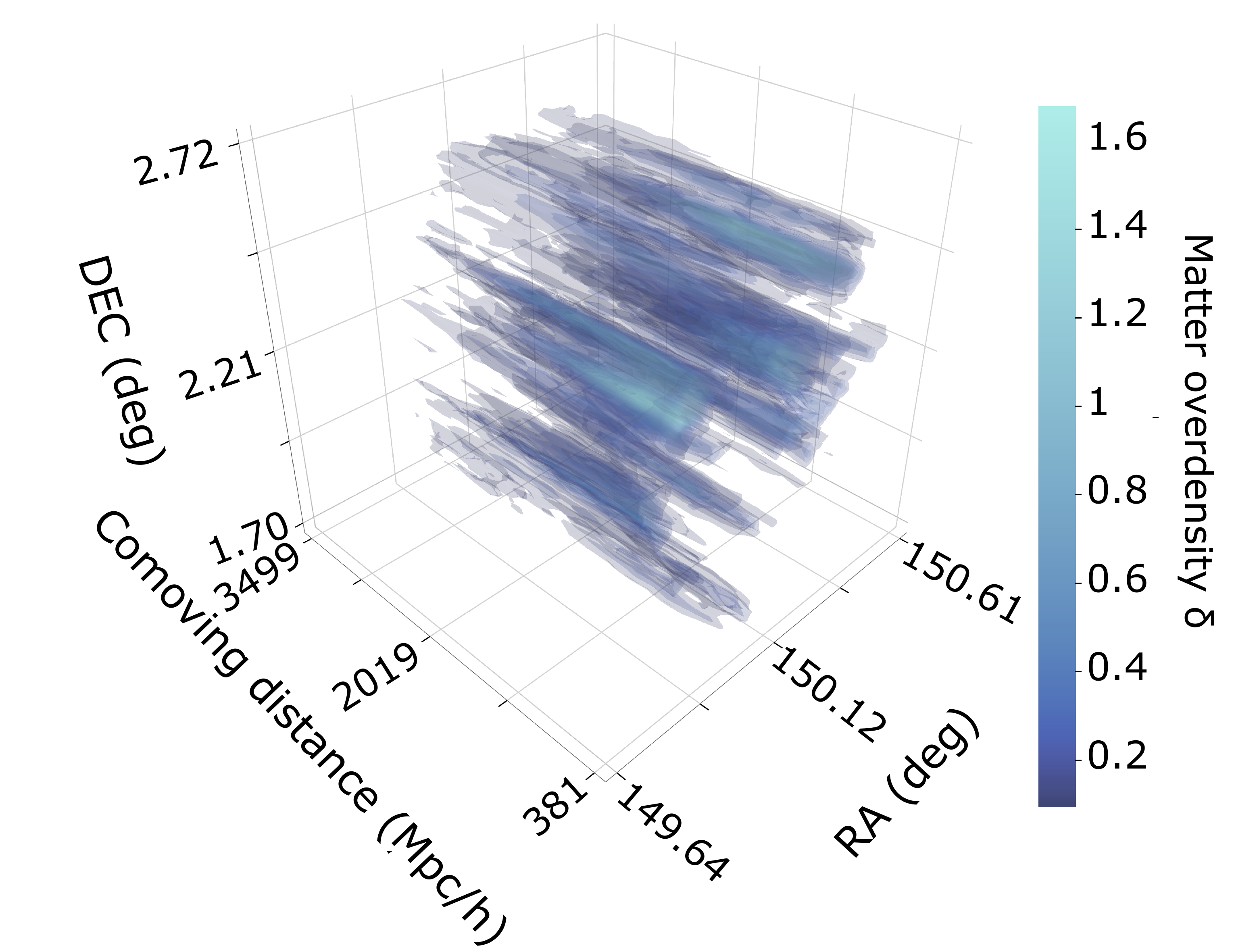}
\end{minipage}

\vspace{0.3em}

% ============================================================
% Row b: Best posterior sample
% ============================================================
\noindent
\begin{minipage}[t]{0.025\textwidth}
\vspace{0pt}
\textsf{\textbf{\large b}}
\end{minipage}%
\hfill
\begin{minipage}[t]{0.965\textwidth}
\vspace{0pt}
\centering
\includegraphics[
    width=0.32\linewidth,
    trim=30 0 30 0,
    clip
]{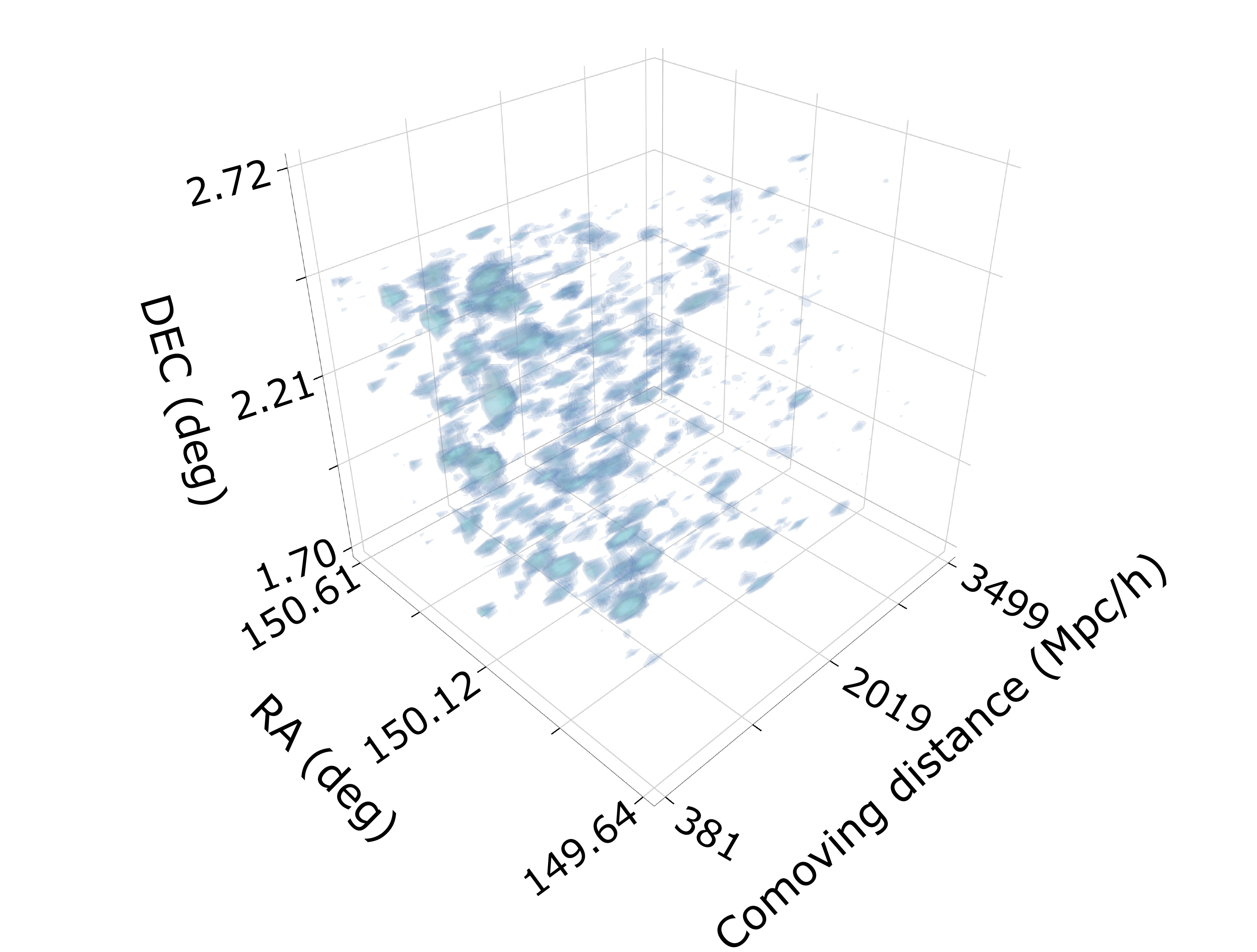}%
\hspace{0.003\linewidth}%
\includegraphics[
    width=0.32\linewidth,
    trim=30 0 30 0,
    clip
]{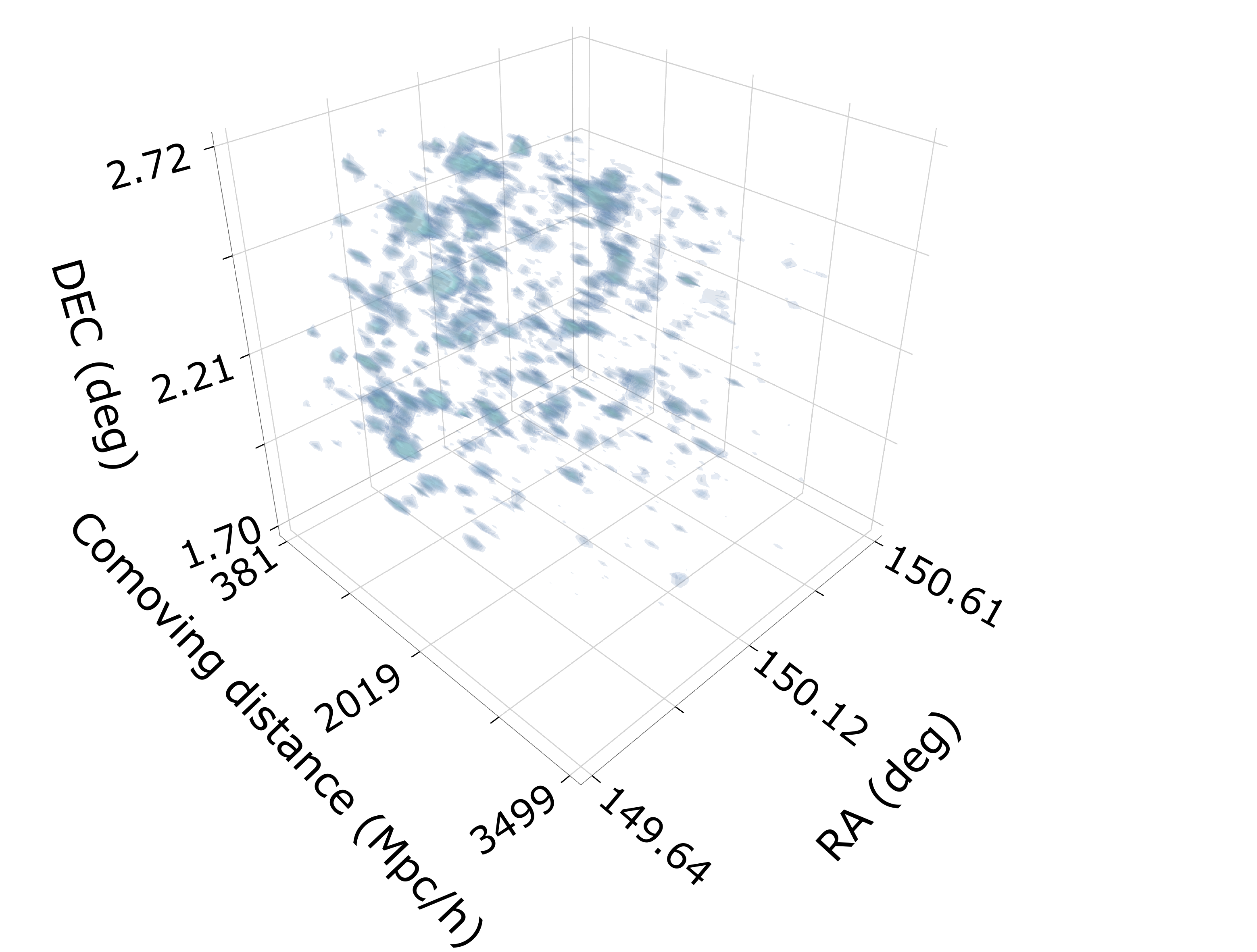}%
\hspace{0.003\linewidth}%
\includegraphics[
    width=0.32\linewidth,
    trim=30 0 30 0,
    clip
]{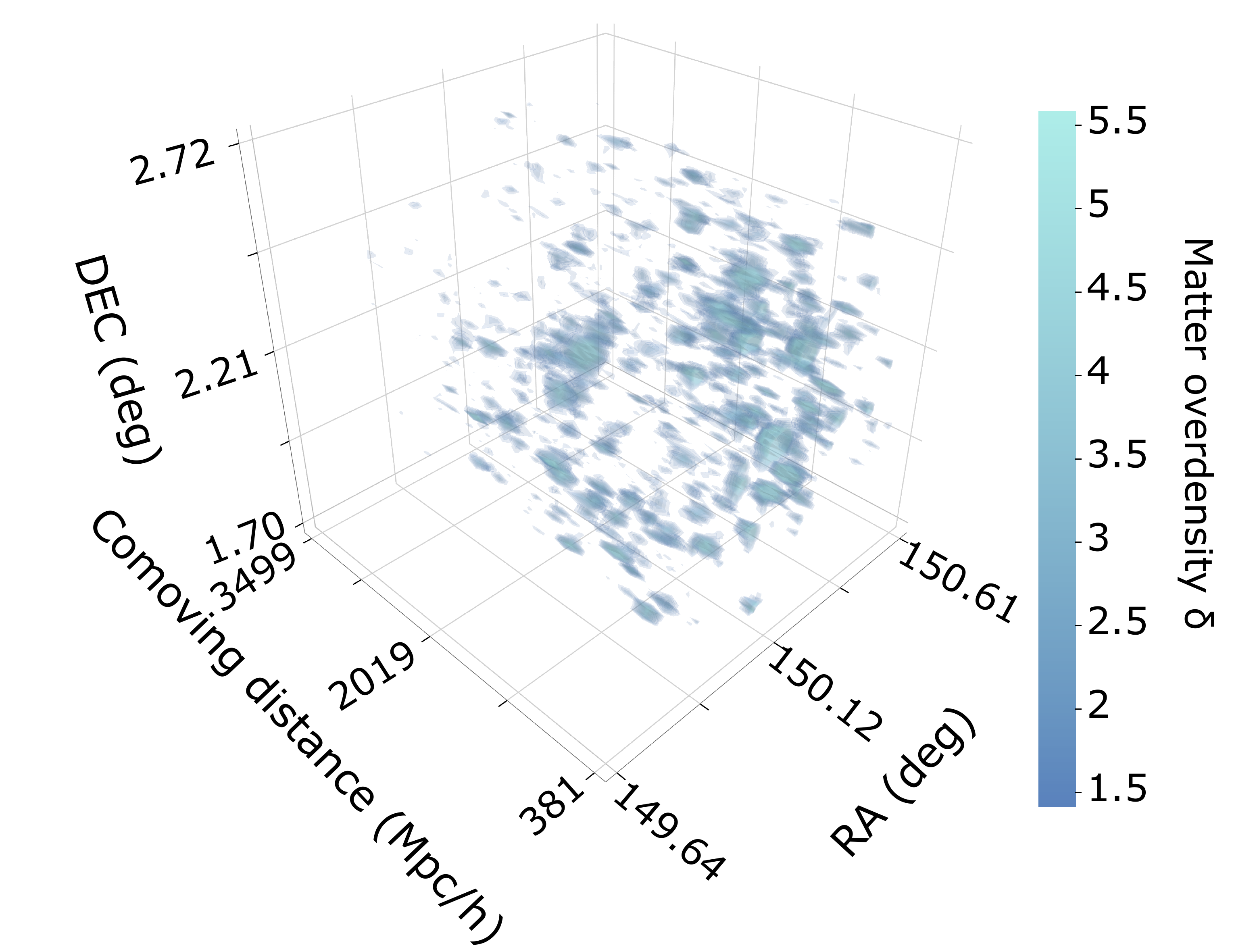}
\end{minipage}

\vspace{0.3em}

% ============================================================
% Row c: Random posterior sample
% ============================================================
\noindent
\begin{minipage}[t]{0.025\textwidth}
\vspace{0pt}
\textsf{\textbf{\large c}}
\end{minipage}%
\hfill
\begin{minipage}[t]{0.965\textwidth}
\vspace{0pt}
\centering
\includegraphics[
    width=0.32\linewidth,
    trim=30 0 30 0,
    clip
]{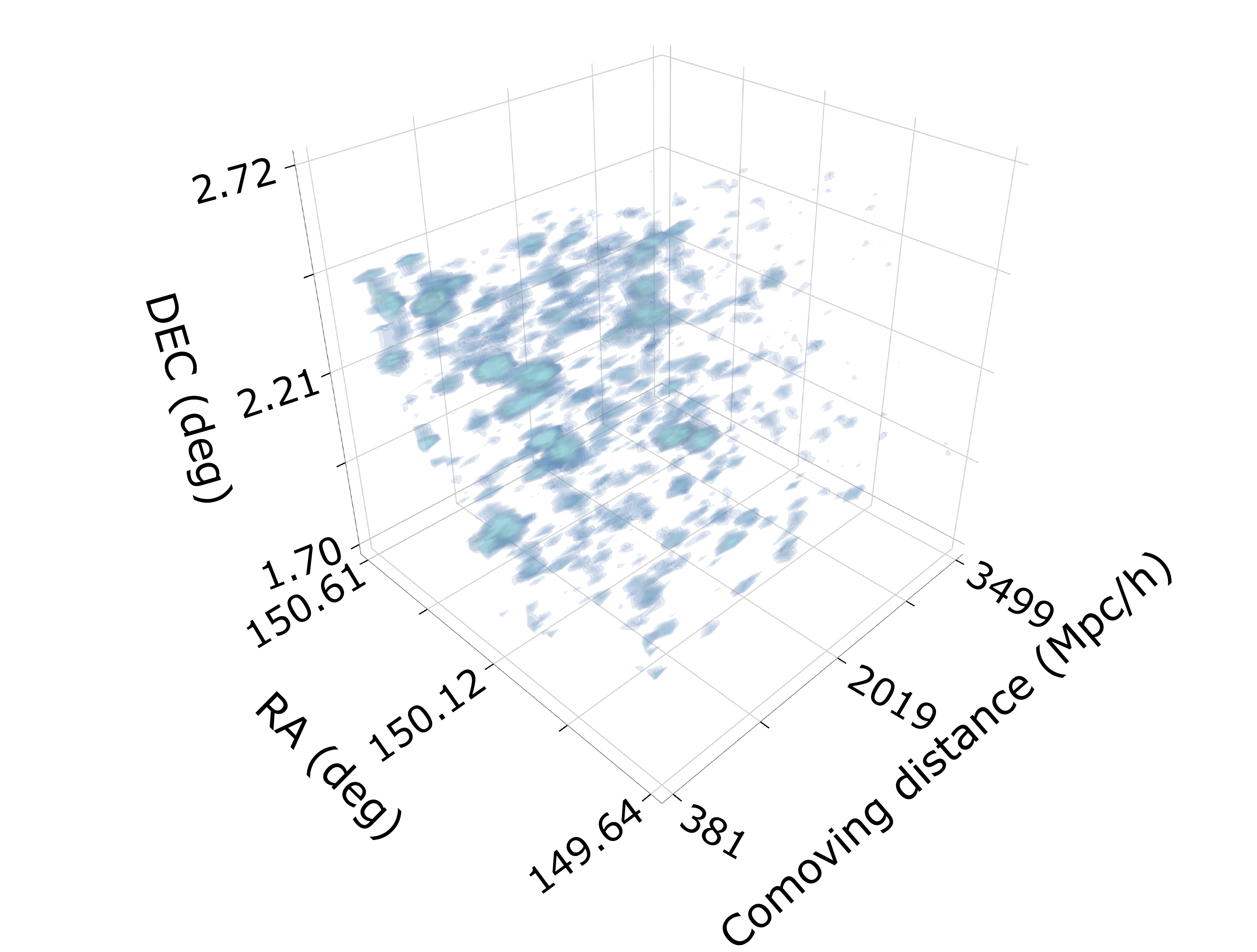}%
\hspace{0.003\linewidth}%
\includegraphics[
    width=0.32\linewidth,
    trim=30 0 30 0,
    clip
]{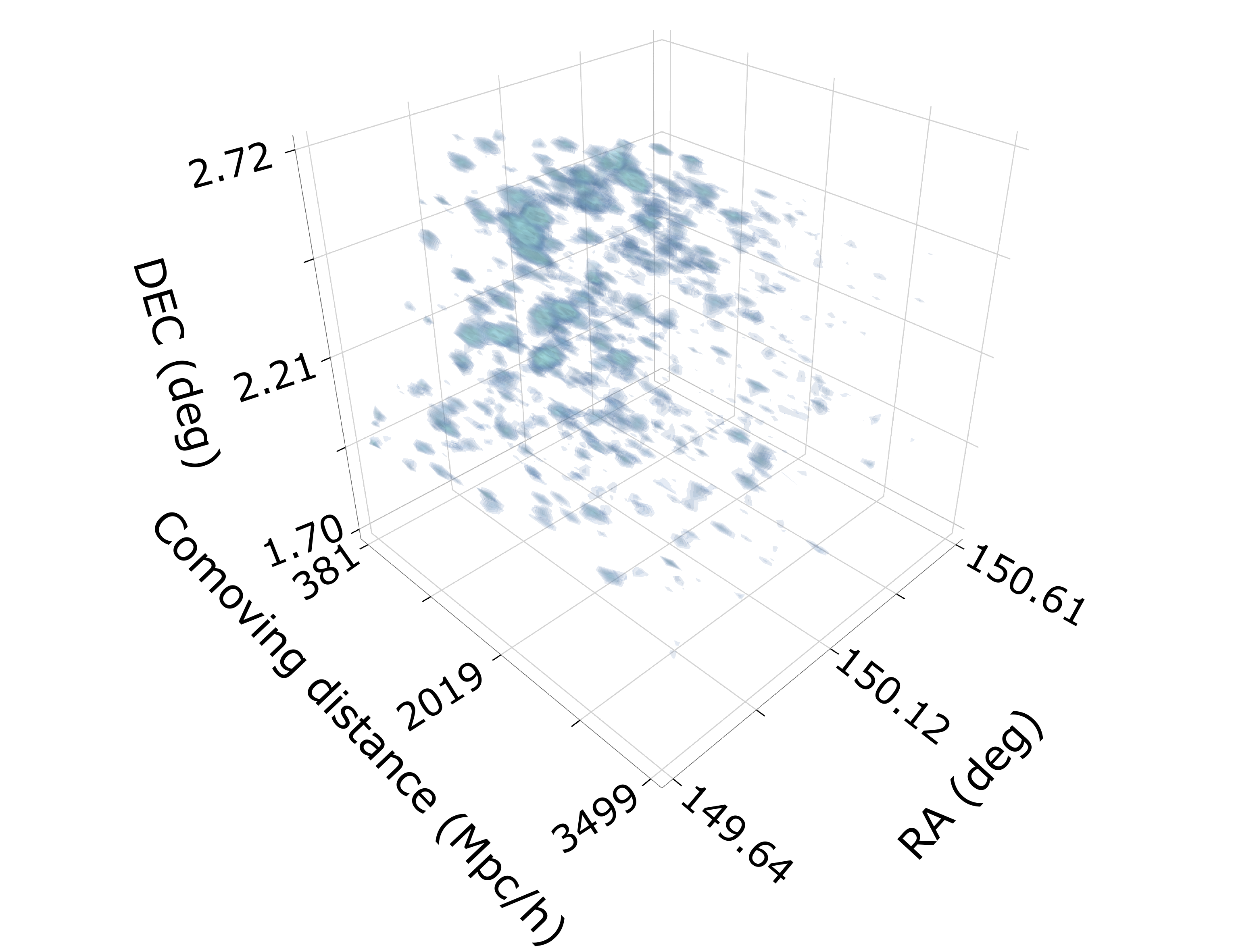}%
\hspace{0.003\linewidth}%
\includegraphics[
    width=0.32\linewidth,
    trim=30 0 30 0,
    clip
]{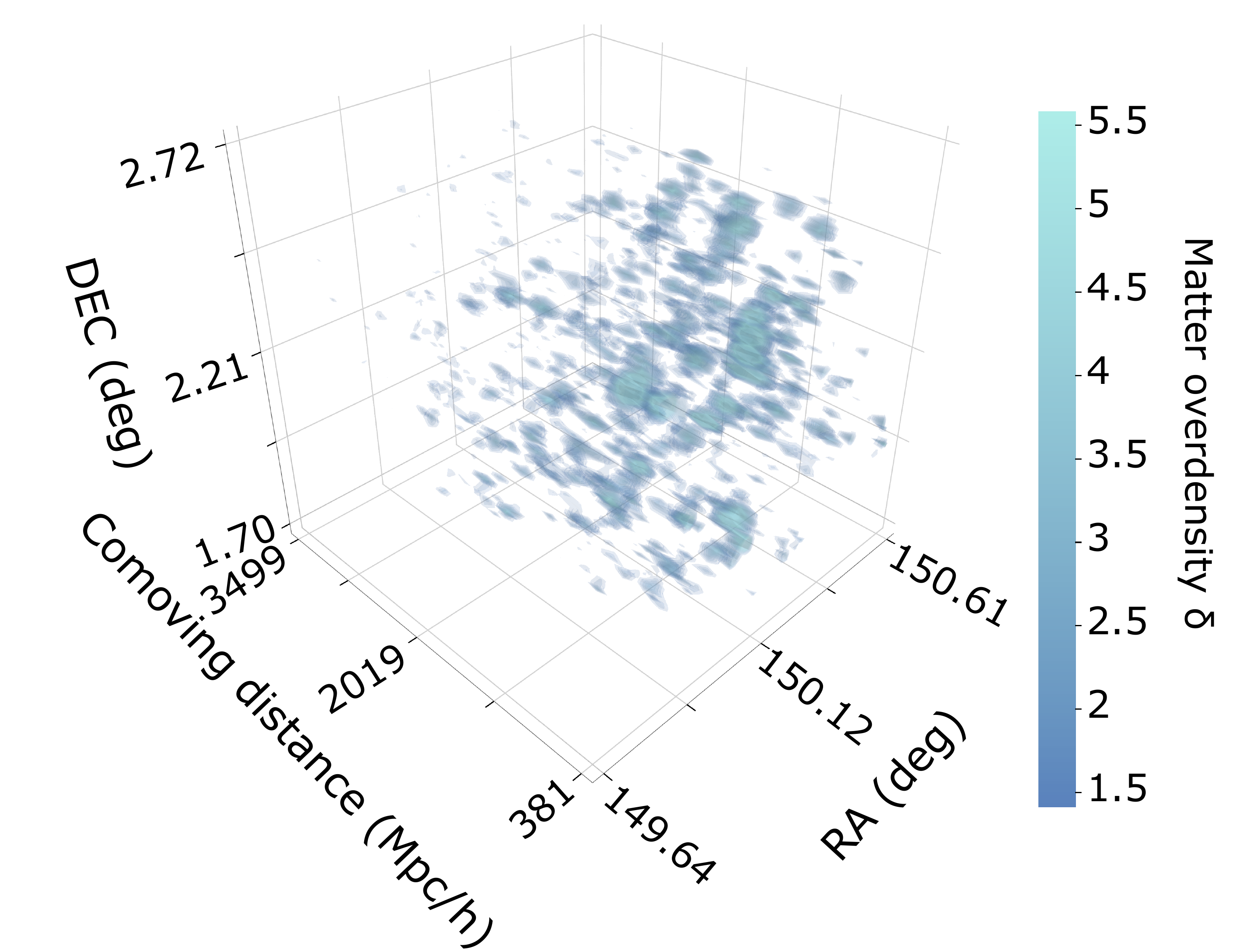}
\end{minipage}

% Caption
% ============================================================
\caption{
\textbf{Three-dimensional posterior reconstruction of the COSMOS-Web matter distribution.}
Three perspective views of the inferred matter overdensity field, $\delta$, across the COSMOS-Web footprint and 20 lens planes spanning $z\simeq 0.13-1.92$. Columns show the same volume from three viewing directions.
\textbf{a.} Posterior mean density field computed from 1,024 posterior realizations, emphasizing structures that remain coherent across the posterior ensemble.
\textbf{b.} Posterior realization with the highest three-dimensional Pearson
correlation with a template constructed from independently identified galaxy clusters \cite{Scognamiglio2026}. This selection is used only for visualization and does not enter the inference or subsequent statistical analyses. \textbf{c.} An unselected posterior realization drawn randomly from the same ensemble, illustrating realization-to-realization variation.
Individual realizations retain sharper, higher-contrast structures than the posterior mean, while variations among them reflect posterior uncertainty in the amplitude, morphology and radial localization of the reconstructed matter distribution. Color indicates matter overdensity; separate color ranges are used for the posterior mean and individual realizations.
}
\label{fig:nine_panels}
\end{figure*}

%%%SLICED WL 3D MASS MAP %%%
\begin{figure*}[h!]
    \centering  \includegraphics[width=\textwidth]{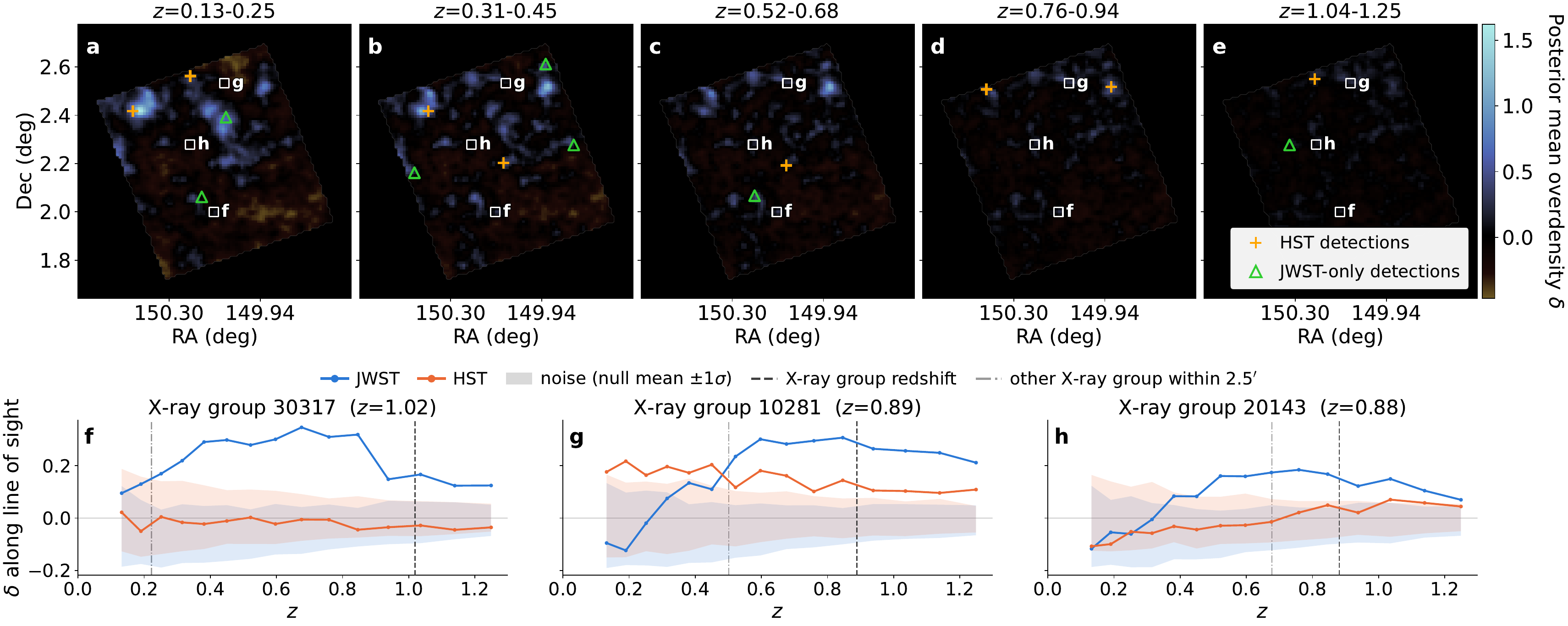}
    \caption{\textbf{Redshift slices of the reconstructed COSMOS-Web matter distribution. a--e.} JWST posterior mean matter overdensity averaged over adjacent lens planes in five redshift intervals ($z=0.13$--$1.25$). Light blue/white and dark brown indicate overdensities and underdensities, respectively. Orange crosses mark galaxy clusters detected in the previous HST weak lensing mass map, whereas green triangles indicate JWST-only detections \cite{Scognamiglio2026}.  Cluster positions and redshifts are independently determined and are not used in the reconstruction. These slices from a common angular footprint highlight the changing location and persistence of matter structures along the line of sight. All slices are displayed at the same angular resolution, corresponding to an increasing transverse physical scale with redshift. White squares (f$-$h) mark the lines of sight examined below. \textbf{f--h.} Posterior mean overdensity profiles from JWST (blue) and HST (orange) towards three X-ray groups from \cite{Gozaliasl2019}. Shaded regions show the mean $\pm1\sigma$ of 128 identically smoothed randomized-shape null reconstructions. Dashed lines mark the selected group redshifts; dash-dotted lines indicate other X-ray groups within 2.5 arcmin. The three examples were selected by inspection to illustrate the JWST–HST contrast.
}  
\label{fig:3d_maps_slices}
\end{figure*}

\noindent
3D WL reconstruction is challenging because the broad lensing kernel mixes contributions from matter at different distances along the line of sight, while intrinsic galaxy-shape noise (arising from the unknown unlensed shapes of individual galaxies) and redshift uncertainties further limit radial information \cite{Heavens2023}. Existing approaches stabilize this ill-posed inversion through simple analytic priors or regularizers \cite{taylor2001,Simon2009,Leonard_2014,Oguri_2017}, which can suppress or restrict the recovery of small-scale, non-Gaussian structure and limit the characterization of uncertainty in the inferred matter field. 
The pioneering \textit{Hubble} Space Telescope (HST) reconstruction of COSMOS demonstrated the feasibility of 3D WL mapping to $z\simeq1$ \cite{Massey2007}, but its source density, depth, and the reconstruction methodology available at the time limited its spatial and radial resolving power, particularly at high redshift.\\

\noindent
Here we combine the depth, source density and angular resolution of JWST COSMOS-Web imaging with generative Bayesian inference to reconstruct the 3D matter distribution across a 0.54 deg$^{2}$ field and 20 lens planes extending to $z\simeq2$. This extends the redshift reach of previous 3D WL reconstructions towards cosmic noon, an epoch of rapid galaxy growth \cite{Madau_2014}. This combination provides access to the 3D matter structure on substantially smaller angular scales than previous wide-field COSMOS WL reconstructions, while retaining the area required to trace the cosmic web rather than individual targeted structures. The WL catalogue used for the reconstruction reaches an effective source density of 121 galaxies arcmin$^{-2}$, enabling detailed three-dimensional mass mapping through dense angular sampling and enhanced sensitivity to matter structures along the line of sight over such a large contiguous extragalactic field. We couple a physical WL forward model to a diffusion-based generative prior learned from cosmological simulations \cite{Zhao_2026}, and sample the resulting high-dimensional posterior distribution of the matter field.\\

\subsection*{3D matter distribution}
\noindent
The reconstructed COSMOS-Web volume reveals matter structure across $z\simeq0.13-2$, retaining non-linear structure across a broad range of spatial scales (Fig.~\ref{fig:nine_panels}). The posterior mean emphasizes structures that remain coherent across the 1,024 posterior realizations, whereas individual samples reveal the range of higher-contrast morphologies permitted by the posterior (Fig.~\ref{fig:nine_panels}a--c).\\

\noindent
The radial structure becomes apparent when the reconstructed volume is separated into redshift
intervals (Fig.~\ref{fig:3d_maps_slices}a--e). The dominant over-densities change in position and amplitude between slices, showing that structures contributing to the projected lensing signal arise at different cosmic distances. Several reconstructed over-densities coincide with independently identified galaxy clusters at consistent angular positions and redshifts. In particular, the JWST reconstruction recovers structures associated with higher-redshift clusters \cite{Scognamiglio2026} that were not detected in the previous HST WL map of COSMOS \cite{Massey2007}, demonstrating the improved sensitivity of JWST-based mass mapping to distant matter structures.\\

\noindent
% The comparison can be extended to X-ray groups outside the reference sample. Fig.~\ref{fig:3d_maps_slices}\textbf{f--h} shows the line of sight through three groups at $z\simeq0.9$--$1.0$ from the full catalogue of \cite{Gozaliasl2019}, chosen by inspection as lines of sight where the JWST reconstruction and a reconstruction of the HST COSMOS shear catalogue \cite{Leauthaud07} with the same pipeline differ clearly (Methods). In each case the JWST posterior mean lies above the range of the randomized-shape null reconstructions over an extended redshift interval that includes the group. Towards 30317 the HST reconstruction remains within its null range at all redshifts; towards 10281 it shows a weaker overdensity spread over a broad redshift range; towards 20143 it stays within or near the edge of its null range. The null ranges of the two reconstructions are comparable, so the difference reflects the stronger lensing response recovered from the denser JWST source sample rather than a difference in noise. These lines of sight are illustrations selected for this contrast, not a complete census.\\
To further examine the recovery of matter structures at high redshift, we compare the JWST reconstruction with the one obtained by applying our proposed inference pipeline to the HST COSMOS shear catalogue of \cite{Leauthaud07}. We consider three independently identified X-ray groups at $z\simeq0.9$--$1.0$ from \cite{Gozaliasl2019}, outside the reference cluster sample, selected by inspection to illustrate differences between the two reconstructions (Fig.~\ref{fig:3d_maps_slices}f--h). At all three positions, the JWST posterior mean shows a more pronounced overdensity relative to its randomized-shape null ensemble, extending across a broad redshift interval encompassing the corresponding X-ray group. In contrast, the HST reconstruction exhibits weaker or less clearly distinguishable responses relative to its own null ensemble. The null scatter is comparable between the two reconstructions, with median JWST-to-HST standard deviation ratios of 0.98, 1.00 and 1.04 across the 15 displayed lens planes, respectively.  This indicates that the stronger JWST response is not simply due to differences in the noise level. The deeper, higher-resolution JWST observations provide a denser sampling of background galaxy shapes, potentially improving sensitivity to matter structures at these redshifts. The reconstructed signal can peak in front of the group redshift, reflecting the limited radial localization inherent to WL. These examples illustrate the additional line-of-sight information recovered from the JWST observations rather than providing a complete census of the X-ray groups.\\

\noindent
The reconstruction should not, however, be interpreted as precisely locating individual structures in redshift. The broad WL efficiency kernel and source-redshift uncertainties couple neighbouring lens planes, introducing uncertainty in the radial positions of individual structures. While individual posterior samples can contain sharply localized structures, their radial positions vary across samples, resulting in broader, correlated features across neighbouring lens planes in the posterior mean. The advance is therefore the ability to distinguish matter structures at different redshifts that overlap in 2D mass maps, while accounting for uncertainties in their radial positions. We quantify this limitation using end-to-end simulations below.\\
% The reconstruction should not, however, be interpreted as assigning individual structures to
% thin redshift planes. The broad weak lensing efficiency kernel and source-redshift
% uncertainties couple neighbouring lens planes, producing correlated radial responses and causing
% some structures to persist across adjacent intervals. The advance is therefore not
% spectroscopic radial resolution, but the recovery of statistically constrained line-of-sight
% information that is absent from two-dimensional weak lensing maps. We quantify the radial
% response and localization of the reconstruction using end-to-end simulations below.\\

\subsection*{Significance and systematics}
%%%2D MASS MAP COMPARISON %%%
\begin{figure*}[h!]
    \centering  \includegraphics[width=\textwidth]{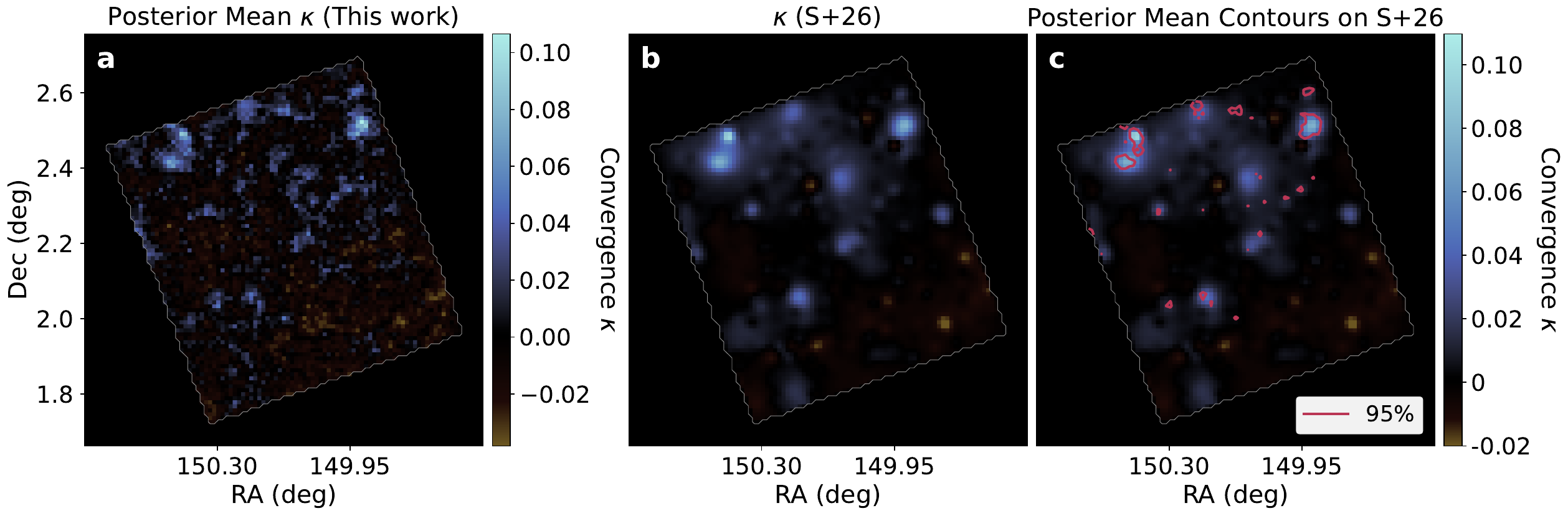}
    \caption{\textbf{Comparison with the two-dimensional COSMOS-Web weak lensing mass map. a.} Posterior mean convergence, $\kappa$, obtained by projecting the inferred three-dimensional matter distribution along the line of sight. \textbf{b.} Previously published COSMOS-Web 2D convergence map (referred as S+26; \cite{Scognamiglio2026}), reconstructed directly from the JWST COSMOS-Web shear catalogue. \textbf{c.} Same two-dimensional convergence map as in \textbf{b.}, overlaid with red contours enclosing regions where the projected three-dimensional posterior mean in \textbf{a.} exceeds the 95th percentile of its positive convergence values. The two reconstructions use the same underlying JWST shear catalogue but distinct inference approaches and the two-dimensional map is not used to constrain the three-dimensional reconstruction. Their spatial correspondence shows that the dominant angular mass concentrations are recovered by both approaches.
} 
    \label{fig:2dmap_comp}
\end{figure*}

\noindent
We first test whether the angular structures recovered in three dimensions are consistent with the independently reconstructed 2D mass map (Fig.~\ref{fig:2dmap_comp}). Projecting the inferred 3D volume back into two dimensions recovers the dominant mass concentrations seen in the COSMOS-Web WL map (referred as S+26; Fig.~\ref{fig:2dmap_comp}b). Although both reconstructions originate from the same JWST shear catalogue, they use distinct inference approaches, and the 2D map is not used to constrain the 3D reconstruction. Their spatial correspondence shows that introducing radial degrees of freedom preserves the angular mass structures recovered directly from the shear catalogue, while the 3D posterior distributes this projected structure along the line of sight.\\
%%%PS COMPARISON %%%
\begin{figure*}[h!]
    \centering  \includegraphics[width=\textwidth]{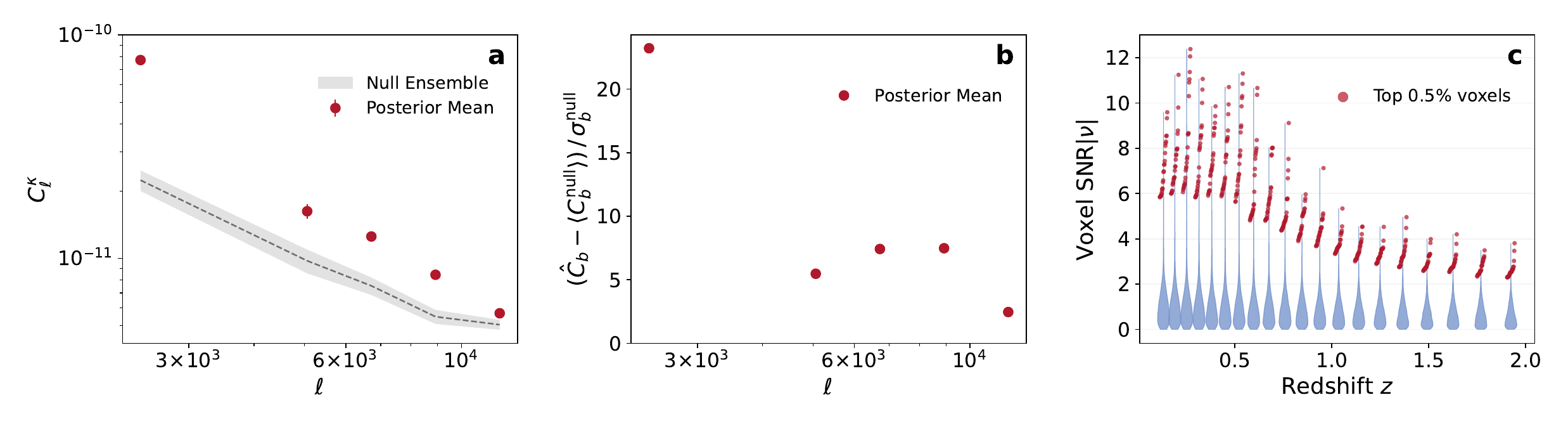} 
    \caption{\textbf{Statistical significance of the reconstructed weak lensing signal. a.} Angular band powers of the projected posterior mean convergence field (red) compared with the mean and 1$\sigma$ scatter of 128 randomized galaxy-shape null reconstructions (grey). The null reconstructions are obtained by independently randomizing the orientations of the observed galaxy ellipticities and rerunning the 3D reconstruction pipeline, thereby removing spatially coherent shear while preserving the measured ellipticity amplitudes and source sampling. Our posterior mean reconstruction exhibits excess angular power relative to the null, indicating a spatially coherent shear signal significantly above the shape-noise level. \textbf{b.} Bandwise signal-to-noise relative to the null ensemble. For each band, this is the difference between the posterior mean band power(red) and the null mean (grey) in \textbf{a}, in units of the null 1$\sigma$ scatter. \textbf{c.} Voxel significance, $|\nu|$, defined as the deviation of the posterior mean from the mean of the null reconstructions in units of their standard deviation at each voxel, as a function of lens-plane redshift. Red points mark the 25 pixels with the largest $|\nu|$ in each plane, corresponding to the upper 0.5\% of in-footprint pixels.}
    \label{fig:PS_compar}
\end{figure*}

% The recovered structure is detected statistically beyond individual visual features. The projected posterior mean contains excess angular power relative to randomized galaxy-shape reconstructions across the resolved multipole range
% (Fig.~\ref{fig:PS_compar}a,b), demonstrating coherent lensing information
% beyond that expected from shape noise.

\noindent
The recovered structure is supported by statistical tests beyond individual visual features. The projected posterior mean exhibits excess angular power relative to reconstructions obtained by randomizing galaxy orientations across the five multipole bands considered  (Fig.~\ref{fig:PS_compar}a, b). 
% These randomized-shape reconstructions preserve the observational sampling and shape-noise properties while removing coherent galaxy orientations, providing an empirical reference for the reconstructed power in the absence of a coherent lensing signal. 
These randomized-shape reconstructions are obtained by independently rotating the observed, already-lensed galaxy ellipticities, rather than their intrinsic unlensed shapes. This destroys the spatial coherence of the measured shear signal while preserving the observed ellipticity amplitudes, galalxy positions, and redshift information, providing an empirical reference for the reconstructed WL power in the absence of coherent lensing. The observed excess therefore indicates that the reconstruction contains angular power beyond that obtained from the shape-noise null ensemble. This comparison does not assess cosmic variance or establish the significance of individual reconstructed structures; rather, it quantifies the excess power relative to the null distribution at each angular scale (Fig.~\ref{fig:PS_compar}b). We next use the same randomized-shape ensemble to assess the local significance of the reconstructed 3D density field. For each voxel, we quantify the significance $\nu$ as the deviation of the posterior mean from the null-ensemble mean in units of the null-ensemble standard deviation (see \nameref{Sim_Val}). The most significant voxels are concentrated predominantly at $z\simeq0.2$--$0.6$, with decreasing significance towards higher redshift (Fig.~\ref{fig:PS_compar}c). This trend reflects the combined effects of lensing efficiency, source sampling, and posterior uncertainty, and identifies the redshift range over which individual structures are most strongly
constrained by the present observations.\\

\noindent
Because the reconstruction is informed by a learned cosmological prior, we further test whether it correctly recovers structures supported by the lensing data rather than simply generating plausible matter configurations. 
%End-to-end reconstructions of 16 independent cosmological light cones withheld from training recover the locations and large-scale morphology of the dominant over-densities and their broad scale-dependent angular structure (Extended Data Fig.~\ref{fig:PS_val}). 
We generate synthetic WL observations from 16 independent cosmological light cones withheld from training, for which the true matter distribution is known. Applying our reconstruction pipeline to these synthetic observations, we recover the locations and large-scale morphology of the dominant over-densities and their broad scale-dependent angular structure (Extended Data Fig.~\ref{fig:PS_val}).  However, the precise radial positions of individual structures remain uncertain. In individual posterior samples, structures can appear sharply localized, but their positions shift between samples. When averaged over the posterior ensemble, these variations produce broader features extending across neighbouring redshift slices in the posterior mean (Extended Data Fig.~\ref{fig:synthetic_radial_recovery}). 
% The radial response is localized around the input redshift intervals but extends into neighbouring slices, as expected from the broad lensing kernel. 
These tests quantify the angular fidelity and radial limitations of the reconstruction and show that the inferred 3D structure is not determined by the generative prior alone.\\

\begin{figure}[h!]
    \centering
\includegraphics[width=\columnwidth]{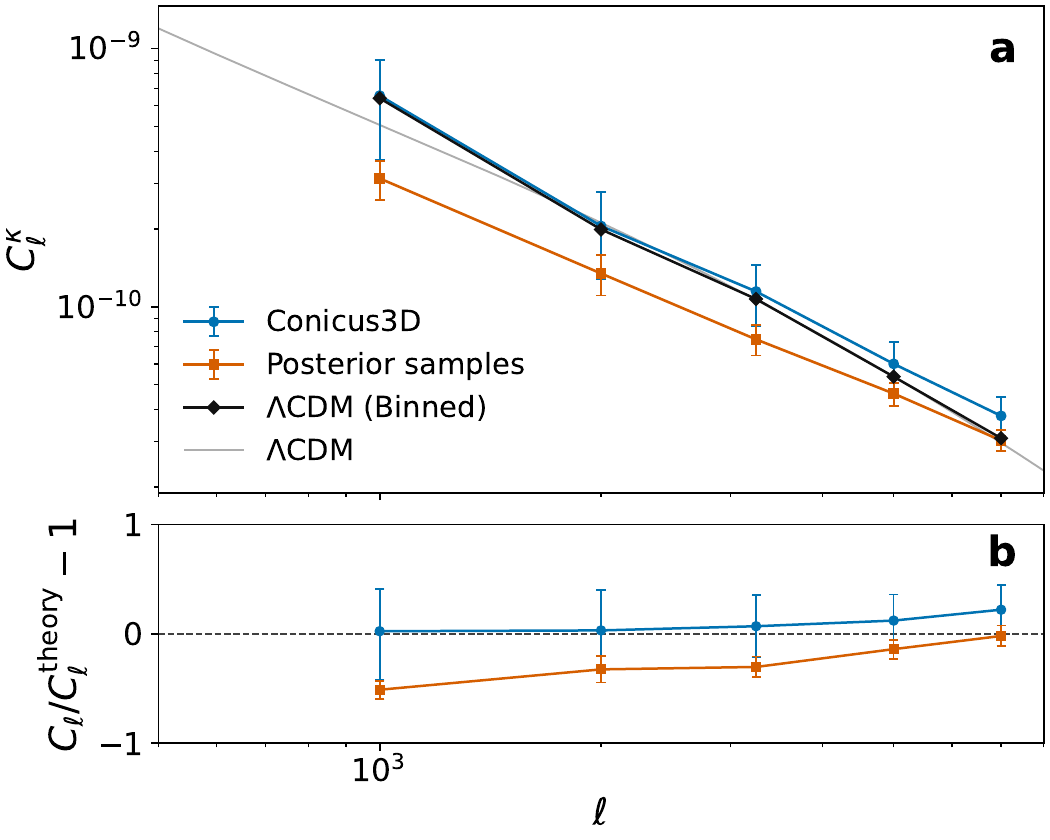}
    \caption{\textbf{Consistency of the reconstructed convergence power with the fiducial \(\Lambda\)CDM cosmology. a.} Angular convergence band powers, $C_{\ell}^{\kappa}$, of the reconstructed COSMOS-Web field and 800 held-out \texttt{Conicus3D} realizations, compared with the fiducial $\Lambda$CDM prediction. Points and error bars show the mean and central 68\% (16th–84th percentile) intervals across 1,024 posterior samples and 800 \texttt{Conicus3D} realizations, respectively. The continuous curve shows the fiducial theoretical spectrum; the binned prediction includes the same survey mask, mode-coupling and band-power operators applied to the reconstructed and simulated fields. \textbf{b.} Fractional band-power residuals of the COSMOS-Web reconstruction and held-out \texttt{Conicus3D} ensemble relative to the binned fiducial prediction. The reconstructed power is lower than the simulated mean on large angular scales but remains statistically consistent with the simulated ensemble when accounting for field-to-field variance and correlations between angular scales.}
    % All fields are projected using the same source-weighted lensing kernel and measured over the same C1-apodized COSMOS-Web footprint using the flat-sky pseudo-$C_{\ell}$ estimator.}
    \label{fig:recon_cosmology}
\end{figure}

\noindent
We also test whether the statistical properties of the reconstructed field are consistent with those expected in the fiducial $\Lambda$CDM cosmology. The angular convergence power spectrum of our COSMOS-Web reconstruction follows the scale dependence of 800 independent simulated matter fields from the \texttt{Conicus3D} dataset (see \nameref{Conicus}) excluded from model training, which provide a reference for the expected variation under the fiducial cosmology. Its mean power lies below the simulated expectation on large angular scales and approaches it towards smaller scales (Fig.~\ref{fig:recon_cosmology}). End-to-end tests on synthetic data show that the reconstruction suppresses the recovered power by $6$--$12\%$ over the scales tested (Extended Data Fig.~\ref{fig:PS_val}d), accounting for part of this difference. Including field-to-field variance and correlations between angular scales,  the reconstructed COSMOS-Web spectrum is statistically consistent with the simulated ensemble ($D^2=1.70$, empirical $p=0.81$; see \nameref{Sim_Val}). Because the generative prior is itself learned from simulations of the fiducial cosmology, this comparison is a consistency test rather than an independent cosmological constraint.\\

%%%RECONSTRUCTED MATTER DISTR> VS PREDICTION%%%

\noindent
Finally, we test whether residual point-spread-function (PSF) anisotropy could produce coherent structure in the reconstruction. We detect no statistically significant leakage of residual stellar PSF ellipticity into the reconstructed shear field (Extended Data Fig.~\ref{fig:PS_val_II}), and no significant scale-dependent correlation between the reconstructed convergence field and PSF-induced convergence maps (Extended Data Fig.~\ref{fig:K_psf_cross_cor}). These PSF tests find no evidence that the recovered structures are produced by residual PSF contamination.\\

\subsection*{Conclusions}
\noindent
Our 3D reconstruction traces the matter distribution across the contiguous 0.54 deg$^{2}$ COSMOS-Web field to $z\simeq2$, extending the redshift reach of previous 3D WL reconstructions towards cosmic noon, while retaining information into the small-scale, non-gaussian regime that encodes the assembly of dark matter haloes and their connection to the surrounding cosmic web. Matter structures that overlap in projection are separated along the line of sight, including overdensities associated with independently identified higher-redshift clusters that were not detected in the previous HST WL map, benefiting from the longer-wavelength coverage, high angular resolution, and sensitivity of JWST observations.\\

\noindent
The generative Bayesian framework further extends the information provided by 3D mass mapping beyond a single reconstructed field. Sampling the high-dimensional posterior allows uncertainty in the amplitude, morphology, and radial localization of matter structures to be quantified and propagated to derived observables. The resulting posterior therefore provides both a high-resolution 3D view of the cosmic web and a probabilistic description of its allowed structure.\\

\noindent
Applied to the much larger WL datasets expected from \textit{Euclid} and the \textit{Nancy Grace Roman Space Telescope}, this approach could trace
the evolution of the cosmic web over substantially larger cosmological volumes and exploit non-Gaussian information in the matter field beyond conventional two-point lensing statistics.

\section*{Methods}
\label{methods}
\subsection*{Observational Data and Redshift Catalogue}
\label{imaging}
The 3D reconstruction uses the JWST COSMOS-Web WL shear catalogue presented in \cite{Scognamiglio2026}. COSMOS-Web (PID 1727; PIs Kartaltepe and Casey) provides contiguous NIRCam imaging over 0.54 deg\(^{2}\) in F115W, F150W, F277W and F444W \cite{Casey_2023}. Galaxy shapes were measured independently in the F115W and F150W images using the \texttt{pyRRG} WL pipeline \cite{Harvey2021}; source detection and multi-band photometry were obtained from the full NIRCam imaging \cite{shuntov2025}. Point-spread-function modelling, source selection, shape measurement and shear calibration follow \cite{Scognamiglio2026}. Combining
the independent F115W and F150W shape measurements yields 251,457 unique WL sources, selected to a limiting magnitude of $\mathrm{MAG_AUTO}<28$ AB, corresponding to an effective source density of 129 galaxies arcmin$^{-2}$.\\

\noindent
Photometric redshifts were taken from the COSMOS2025 catalogue
\cite{shuntov2025}, derived using the \texttt{LePhare} template-fitting
framework \cite{Ilbert_2006}, and supplemented with secure spectroscopic redshifts from the \texttt{cosmosastro} compilation \cite{khostovan2025}. The WL and spectroscopic catalogues were cross-matched within
0.1 arcsec. We retain only spectroscopic measurements with quality flags corresponding to greater than 99\% confidence in the redshift identification,
yielding 14,603 WL sources with secure spectroscopic redshifts.
For these sources, the spectroscopic measurement replaces the photometric
estimate; all remaining sources retain their photometric-redshift information.\\

\noindent
For the 3D reconstruction, we retain sources with secure spectroscopic redshifts satisfying $0.131<z_{\rm spec}\leq15$ and sources
with photometric redshifts satisfying $10^{-4}<z_{\rm phot}\leq15$ \cite{khostovan2025}. 
The lower photometric-redshift threshold is more permissive because photometric-redshift uncertainties are propagated through the lensing forward model, allowing sources with best-fitting redshifts below the first lens plane to retain non-zero probability behind that plane; spectroscopic sources below
the first lens plane are instead excluded. 
These selections yield 235,911 sources, corresponding to an effective source density of
121 galaxies arcmin$^{-2}$. The spectroscopic and photometric-redshift information enters the observational forward model described below and is not
incorporated into the data-driven prior.\\

\noindent
For independent validation of the radial reconstruction, we use a reference sample of 15 galaxy clusters identified within the COSMOS-Web footprint through extended X-ray emission detected with XMM-\textit{Newton} and
\textit{Chandra} \cite{Gozaliasl2019}. These systems were cross-matched with the optical cluster catalogue of \cite{Toni2025} using a projected physical
separation of $0.5\,{\rm Mpc}\,h^{-1}$ and a redshift tolerance of
$\Delta z=0.05(1+z)$. When multiple associations were present, optical
systems were prioritized according to their detection significance and X-ray systems according to their $M_{200}$ halo mass \cite{Scognamiglio2026}. The resulting reference catalogue provides independent sky positions,
redshifts and $M_{200}$ estimates for comparison with the 3D WL reconstruction.\\

\subsection*{3D Reconstruction Framework}
\label{sec:recon_framework}
We reconstruct the 3D matter distribution using the method introduced in our previous work \cite{Zhao_2026}, applied here to the JWST COSMOS-Web observations. We infer the posterior distribution of the 3D matter overdensity field, $\delta$, conditioned on the observed JWST shear measurements, $\gamma$,\\
\begin{equation}
\label{eq:posterior}
p(\delta|\gamma)\propto p(\gamma|\delta)\,p(\delta),
\end{equation}\,

\noindent
where the prior, $p(\delta)$, is a diffusion model learned from the \texttt{Conicus3D} simulations (see \nameref{Conicus}) and the likelihood, $p(\gamma|\delta)$, is defined by the WL forward model and the shape-noise covariance of the observations (see \nameref{sec:WL_formalism}). The likelihood constrains the reconstruction to matter fields supported by the observed shear, while the prior regularizes spatial structure that is weakly constrained by the data. Posterior samples are drawn by combining the two terms in an inference approach that makes use of reverse diffusion (see \nameref{diffusion}).

\subsection*{Conicus3D Simulations and Diffusion Prior}
\label{Conicus}
\subsubsection*{Simulation dataset}
We use the \texttt{Conicus3D} dataset \cite{Zhao_2026} to construct a data-driven prior, $p(\delta)$, for the 3D matter distribution. \texttt{Conicus3D} consists of dark matter light cones constructed from the
AbacusSummit $N$-body simulations \cite{hadzhiyska2022halo} in the
\texttt{c000} fiducial $\Lambda$CDM cosmology, with
$\Omega_{\rm m}=0.3152$, $\Omega_{\rm b}=0.0493$, $h=0.6736$, and $\sigma_8=0.8080$. Each light cone comprises 20 matter overdensity lens planes equally spaced in comoving distance, with effective redshifts spanning
$z=0.13$--$1.92$. Each plane covers $1.36^\circ\times1.36^\circ$ on a
$128\times128$ grid, corresponding to an angular pixel scale of 38.25 arcsec. We adopt the same lens-plane geometry for the COSMOS-Web reconstruction.
The prior model is trained on phases \texttt{ph001}--\texttt{ph024} of the \texttt{Conicus3D} dataset, which are independent realizations of the simulation initial conditions, comprising 19,200 3D light cones, each represented by 20 lens planes.
%or 384,000 individual lens-plane maps. 
Phase \texttt{ph000} is reserved for held-out evaluation, and no maps from this phase are used during model training.\\ 
%Extended Data Fig.~\ref{fig:PS_val} evaluates four held-out light cones from \texttt{ph000\_lc1}.\\

\noindent
\subsubsection*{Diffusion model architecture and training}
\label{diffusion}
Rather than imposing an analytic Gaussian prior or explicit spatial smoothing, we learn the redshift-dependent distribution of matter density fields from the \texttt{Conicus3D} simulations using the diffusion-based generative framework introduced in \cite{Zhao_2026}. A diffusion model defines a forward process that progressively perturbs the simulated lens-plane density fields with Gaussian noise over a diffusion time $t\in[0,1]$, from the clean field at $t=0$ to pure noise at $t=1$, and is trained to reverse this process, thereby capturing the non-Gaussian spatial statistics of the matter distribution within each lens plane. In particular, we train a neural network to learn the score function\\
\begin{equation}
s_\phi(\delta_t,t)= \nabla_{\delta_t}
\log p_t(\delta_t),
\end{equation}\,

\noindent
where $\delta_{t}$ denotes the noise-perturbed density field at diffusion time $t$, $p_{t}(\delta_{t})$ is its probability distribution at that noise level, and $s_{\phi}$ is the neural-network approximation to its score. 
%the gradient of its log probability density. 
The learned score provides the direction in density-field space towards configurations that are increasingly probable under the diffusion model.\\

\noindent
We parameterize the score of the diffusion model using a U-Net \cite{ronneberger2015u} neural network in the DDPM++ configuration of \cite{song2021score} operating on individual $128\times128$ lens-plane maps and conditioned on lens-plane index. Lens planes are generated independently by the diffusion model; at the radial separations considered here, correlations between lens planes are negligible in the underlying simulations. The network has a base channel width of 128, channel multipliers of 1, 2 and 3, one residual block per resolution level and a self-attention bottleneck. We use a variance-preserving diffusion objective with $\beta_{\min}=0.1$ and $\beta_{\mathrm{d}}=19.9$, sampling continuous diffusion time uniformly over $t\in[10^{-5},1]$ and weighting the denoising loss by the inverse noise variance. Training uses the Adam optimizer \cite{Kingma2017} for
10,000 steps with a batch size of 128 and a learning rate of $10^{-4}$,
including a 500-step linear learning-rate warm-up and an exponential moving average of the network weights.\\
% Optimization uses Adam for 10,000 steps with a batch size of 128 and a learning rate of $10^{-4}$, including linear learning-rate warm-up over the first 500 steps and an exponential moving average of the network weights with a target half-life of 209,600 training examples and a ramp-up ratio of 0.05 \cite{Zhao_2026}.\\

\noindent
During reconstruction, the trained diffusion model provides the prior component of the Bayesian inference, whereas the WL likelihood, defined in the following section (see \nameref{sec:WL_formalism}), determines which matter configurations are supported by the observed JWST shear measurements. The diffusion model therefore does not predict the COSMOS-Web density field directly: it defines a prior distribution over matter configurations learned from the simulations, which is then updated by the observational likelihood. Neither the observed galaxy shears nor their photometric or spectroscopic redshift information are used to train the prior.\\

\noindent
Our diffusion prior is trained on dark-matter-only (N)-body simulations and therefore does not explicitly account for baryonic processes, which can modify the matter distribution on non-linear scales \cite{Schneider2015, Schneider2019}. The lensing likelihood, however, responds to the total matter distribution, including baryons. Baryonic effects should therefore enter mainly on the smallest, noise-dominated scales, where the reconstruction is most prior-dependent and where structure is largely averaged out in the posterior mean.

\subsection*{3D Weak Lensing Inference}
\label{sec:WL_formalism}
\subsubsection*{Forward model and likelihood}
Our goal in 3D reconstruction is to infer the matter overdensity field, $\delta$, from the observed WL shear field, $\gamma$, by modelling the lensing response along the line of sight. The matter distribution is discretized into the 20 lens planes defined by the \texttt{Conicus3D} geometry. Source galaxies are grouped into 30 redshift bins whose lower boundaries follow the lens-plane spacing in comoving distance; the highest-redshift bin includes all sources above the final boundary, with the input catalogue extending to $z=15$.\\

\noindent
For source bin $s$, the convergence at angular position $\boldsymbol{\theta}$ is modelled as\\
\begin{equation}
\label{eq:kappa_discrete}
\kappa_s(\boldsymbol\theta) = \sum_{i} W_{s,i}\,\delta_i(\boldsymbol\theta),
\end{equation}\,

\noindent
where $\delta_{i}$ 
is the matter overdensity in lens plane $i$ and $W_{s,i}$ is the lensing efficiency matrix describing the contribution of that lens plane to sources in bin $s$. The weights are obtained by integrating the geometric lensing kernel over the source-redshift distribution of each bin. Photometric-redshift uncertainties are propagated through the forward model using a Gaussian redshift-error distribution with width $\sigma_z=0.11(1+z)$, whereas sources with secure spectroscopic redshifts are treated at their measured redshift values.\\

\noindent
In the WL limit, we approximate the reduced shear inferred from galaxy ellipticities as $g=\gamma/(1-\kappa)\simeq\gamma$. The shear and convergence fields are related in Fourier space through the Kaiser–Squires relation \cite{Kaiser1993},
\begin{equation}
\label{eq:KS}
\hat{\gamma}(\boldsymbol\ell) = D(\boldsymbol\ell)\,\hat{\kappa}(\boldsymbol\ell)\,
\end{equation}
where\\
\begin{equation}
\label{eq:KS_2}
D(\boldsymbol\ell) = \frac{\ell_1^2 - \ell_2^2 + 2i\,\ell_1\ell_2}{|\boldsymbol\ell|^2}\,
\end{equation}
$\boldsymbol{\ell}$ is the angular wavevector and hats denote Fourier transforms. The complete forward model can therefore be written as a linear operator $F\equiv\mathbf{P}\mathbf{Q}$, where $\mathbf{Q}$ projects the 3D density field onto source-bin convergence maps (Eq.~\eqref{eq:kappa_discrete}) and $\mathbf{P}$ transforms convergence into shear (Eqs.~\eqref{eq:KS} and \eqref{eq:KS_2}).\\

\noindent
This operator defines the likelihood term of the posterior in Eq.~\eqref{eq:posterior}. Assuming Gaussian residuals for the pixel-binned mean shear measurements, the likelihood is\\
\begin{equation}
p(\gamma|\delta)
\propto
\exp
\left[
-\frac{1}{2}
\left(
\gamma-F(\delta)
\right)^T
N^{-1}
\left(
\gamma-F(\delta)
\right)
\right],
\end{equation}\,

\noindent
where $F(\delta)$ is the WL forward model and $N$ is the diagonal covariance of the pixel-averaged shear, with variance $\sigma_s^2/n_s(\boldsymbol\theta)$ in pixel $\boldsymbol\theta$ of source bin $s$, where $n_s(\boldsymbol\theta)$ is the number of sources in that pixel and $\sigma_s$ is the empirical per-component scatter of the measured shear over all sources in bin $s$.\\

\noindent
\subsubsection*{Posterior sampling and visualization}
Posterior samples are generated using the simulation-based Bayesian inference
framework of \cite{Zhao_2026}. At each noise level, the sampler combines a
reverse-diffusion update under the learned prior with a linear-Gaussian
conditioning step incorporating the WL likelihood. The resulting posterior score is\\
\begin{equation}
\nabla_\delta
\log p(\delta|\gamma)= \nabla_\delta
\log p(\gamma|\delta)
+
\nabla_\delta
\log p(\delta).
\end{equation}\,

\noindent
We generate 1,024 posterior realizations of the COSMOS-Web matter field. Rather than yielding a single deterministic reconstruction, the inference therefore returns a posterior distribution of 3D matter fields consistent with the WL observations and the simulation-trained prior. This enables uncertainties in the amplitude, morphology and radial localization of reconstructed structures to be quantified directly from the posterior ensemble. The posterior mean highlights structures that remain coherent across the ensemble, whereas individual realizations retain the range of matter configurations permitted by the posterior (Fig.~\ref{fig:nine_panels}). \\

\noindent
The posterior mean is computed voxel-by-voxel across all 1,024 posterior
realizations and summarizes the ensemble-averaged density field
(Fig.~\ref{fig:nine_panels}a). To illustrate the range of 3D structures represented within the posterior, we also show a realization selected because independently identified massive structures are strongly represented in it, and a realization drawn randomly from the ensemble (Fig.~\ref{fig:nine_panels}). We construct a sparse 3D template from the independent galaxy cluster catalogue described above. Each cluster is assigned
to the lens plane and angular pixel corresponding to its redshift and sky
position, with the corresponding voxel weighted by its $M_{200}$ in units of
$10^{13}\,M_\odot$. If multiple clusters occupy the same voxel, their masses are summed; clusters outside the reconstructed volume or survey footprint are excluded. We compute the 3D Pearson correlation coefficient between this template and each posterior realization, restricted to the non-zero template voxels and without smoothing, thresholding or other preprocessing. The realization with the highest finite correlation coefficient has $r=0.935$ (sample 111, Fig.~\ref{fig:nine_panels}b). This selection is used solely for illustration and does not enter the inference, constitute an independent validation statistic, or contribute to any subsequent statistical analysis. The randomly selected realization (sample 91; random seed 42; Fig.~\ref{fig:nine_panels}c) provides a complementary illustration
of realization-to-realization variation.\\

\subsection*{Reconstruction Validation}
\label{Sim_Val}
We assess the statistical significance, reconstruction fidelity and cosmological consistency of the inferred matter field using complementary data- and simulation-based tests. Randomized galaxy-shape realizations quantify the structure expected in the absence of a coherent lensing signal; end-to-end reconstructions of held-out \texttt{Conicus3D} light cones test the recovery of known matter fields under realistic observational conditions; and an independent ensemble of held-out simulated fields provides the reference distribution for assessing the cosmological consistency of the angular statistics of the COSMOS-Web reconstruction.\\

\noindent
\subsubsection*{Observational null tests}
To construct the observational null distribution, we generate 128 randomized-shape catalogues by independently rotating the measured ellipticity of each source, $e_1+ie_2\rightarrow(e_1+ie_2)\,e^{2i\varphi}$, by an angle $\varphi$ drawn uniformly over $[0,\pi)$. Because the ellipticity is a spin-2 quantity, this rotation spans the full range of orientations. Crucially, the rotation is applied to the observed ellipticities, which already include the lensing-induced distortion, rather than to the intrinsic unlensed galaxy shapes. Rotating intrinsic shapes before lensing would preserve the coherent shear signal, whereas independently rotating the observed ellipticities removes the coherent lensing signal in the ensemble average. This procedure preserves the measured ellipticity amplitudes, source positions and redshift information while randomizing the observed galaxy orientations. Each randomized catalogue is processed through the same 3D inference pipeline as the observations, generating 128 posterior realizations from which we compute a
posterior mean null field. The resulting ensemble of 128 null posterior means propagates shape noise through the full inference and provides an empirical reference for structures recovered in the absence of coherent lensing signal.\\

\noindent
We first quantify the statistical significance of the projected reconstruction
through its angular convergence power spectrum. For the observed field and
each randomized-shape null, we project the 3D posterior mean
into convergence and estimate the angular band powers using the flat-sky,
spin-0 pseudo-$C_{\ell}$ estimator implemented in \texttt{NaMaster}
\cite{Alonso2019}. We apply a 3-arcmin C1 apodization to the survey mask and correct the band powers for mask-induced mode coupling and the pixel window. We use five multipole bands spanning $500 \leq \ell \leq 13{,}500$, with edges at $\ell = 500$, 4,347, 5,771, 7,661, 10,170, and 13,500. For each band $b$, we define the signal-to-noise ratio relative to the randomized-shape ensemble as\\
\[
\mathrm{S/N}_b =
\frac{
\hat{C}_b^{\kappa} - \left\langle \hat{C}_b^{\kappa,\mathrm{null}} \right\rangle
}{
\sigma_b^{\mathrm{null}}
}\,
\]
\noindent
where $\hat{C}_b^{\kappa}$ 
is the measured convergence band power, $\left\langle \hat{C}_b^{\kappa,\mathrm{null}} \right\rangle$ is the mean band power across the 128 null reconstructions, and $\sigma_b^{\mathrm{null}}$ their standard deviation. This statistic measures the excess reconstructed power in units of the empirical null-ensemble scatter (Fig.~\ref{fig:PS_compar}a, b).\\

\noindent
We next use the randomized-shape ensemble to quantify the local significance of the 3D density field. For lens plane $i$ and angular position $\boldsymbol{\theta}$, we define the null-standardized statistic
\[
\nu_i(\boldsymbol\theta) = \frac{\bar{\delta}_{i,\mathrm{data}}(\boldsymbol\theta)-\mu_{i,\mathrm{null}}(\boldsymbol\theta)}{s_{i,\mathrm{null}}(\boldsymbol\theta)}.
\]
Here $\bar{\delta}_{i,\mathrm{data}}$ is the posterior mean computed from the
1,024 COSMOS-Web realizations, whereas $\mu_{i,\mathrm{null}}$ and $s_{i,\mathrm{null}}$ are respectively the mean and sample standard deviation
of the 128 null posterior-mean fields at each voxel. We subtract the angular
mean of each lens plane before computing $\nu$. Within each plane, we rank all finite pixels inside the survey footprint by $|\nu|$; the 25 highest-ranked pixels correspond to the upper 0.5\% of the 4,888 in-footprint pixels (Fig.~\ref{fig:PS_compar}c). The most significant voxels lie predominantly at $z\simeq0.2$--$0.6$, and their significance decreases towards higher redshift, identifying the redshift range over which individual structures are most strongly constrained by the data.\\

\noindent
\subsubsection*{End-to-end simulation tests}
We independently test reconstruction fidelity using 16 randomly selected
\texttt{Conicus3D} light cones from the held-out phase \texttt{ph000}, which is excluded entirely from training. For each held-out field, we generate a synthetic WL catalogue using the positions and redshift information of the COSMOS-Web sources, thereby preserving the survey
footprint, source sampling and redshift distribution of the data. Shear is generated from the known matter field at each source position. Sources with secure spectroscopy retain their measured redshifts, whereas latent true redshifts for photometric sources are drawn from the same Gaussian redshift-error model adopted in the inference, $\sigma_z=0.11(1+z)$. Shape noise is introduced by randomizing the orientation of each observed COSMOS-Web galaxy ellipticity while preserving its amplitude. The synthetic catalogues are analysed using the same source binning,
photometric-redshift marginalization, empirical noise covariance and posterior sampling procedure as the observations. These end-to-end tests therefore propagate realistic source sampling, redshift uncertainty and shape noise through the complete inference while retaining exact knowledge of the input matter field.\\

\noindent
We evaluate recovery of the projected matter field in both configuration and
Fourier space. For each validation field, the known 3D density field is projected into a ground-truth convergence map using the same lensing projection adopted in the inference and compared with the posterior-mean convergence reconstructed from the corresponding synthetic observation. At the native 0.6375-arcmin pixel scale, the median masked Pearson correlation
across the 16 fields is 0.70 and the median root-mean-square error (RMSE) is
0.0199 (Extended Data Fig.~\ref{fig:PS_val}a--c). To test whether the reconstruction benefits from additional angular smoothing,
we apply Gaussian smoothing with standard deviation $\theta=0$--$0.5$ arcmin, in steps of 0.1 arcmin, to the
posterior-mean maps while retaining the ground-truth maps at their native
resolution. The correlation remains approximately 0.70 for $\theta\leq0.2$ arcmin and decreases to 0.63 at $\theta=0.5$ arcmin, while the RMSE increases from 0.0199 to 0.0221. Additional smoothing therefore does not improve agreement with the known projected matter field over the range
tested.\\

\noindent
We quantify the recovery of scale-dependent convergence power through the
fractional band-power residual\\
\begin{equation}
    \Delta_b^C = \frac{\left\langle C_b^{\kappa,\rm{rec}} \right\rangle-C_b^{\kappa,\rm{truth}}}
{C_b^{\kappa,\rm{truth}}}
    = \frac{\left\langle C_b^{\kappa,\rm{rec}} \right\rangle}
{C_b^{\kappa,\rm{truth}}}-1\,
\end{equation}
\noindent
where ${\left\langle C_b^{\kappa,\rm{rec}} \right\rangle}$  is the mean band power across the 1,024 posterior samples of each synthetic observation and $C_b^{\kappa,\rm{truth}}$
is the band power of its corresponding projected ground-truth field. This statistic directly measures reconstruction-induced power bias because each reconstructed field is compared with its own known input realization. The analogous fractional statistic used for the COSMOS-Web cosmological comparison instead takes the binned fiducial $\Lambda$CDM prediction as its reference (Fig.~\ref{fig:recon_cosmology}b). Both quantities are measured using the same lensing projection, survey mask, apodization, mode-coupling correction and band-power operator. Across the five multipole bands, the median fractional band-power residual over the 16 validation fields ranges from $-0.12$ to $-0.06$ (Extended Data Fig.~\ref{fig:PS_val}d), corresponding to a 6--12\% suppression of recovered convergence power over the scales tested. Comparable suppression of recovered signal power has been observed in simulation tests of other regularized WL mass-mapping methods, where priors and smoothing reduce power in noise-dominated modes \cite{des_maps}. The reconstruction therefore recovers the broad scale dependence of the projected matter power while modestly suppressing its amplitude.\\

\noindent
We additionally quantify the Fourier-space correspondence between each posterior mean reconstruction and its matched ground-truth field using the normalized cross-power\\
\begin{equation}
r_\ell =
\frac{C_\ell^{\kappa^{\rm truth}\times\kappa^{\rm post}}}
{\sqrt{C_\ell^{\kappa^{\rm truth}\times\kappa^{\rm truth}}
C_\ell^{\kappa^{\rm post}\times\kappa^{\rm post}}}}\,,
\end{equation}
\noindent
where $\kappa^{\rm post}$ denotes the posterior mean convergence map and the spectra are band-averaged at each effective multipole. Whereas the fractional band-power residual measures recovery of the signal amplitude, $r_\ell$ measures the scale-dependent correspondence between the reconstructed and ground-truth fields after normalization by their respective auto-power. The median $r_\ell$ decreases from 0.92 at $\ell=1{,}000$ to 0.69 at $\ell=7{,}000$ (Extended Data Fig.~\ref{fig:PS_val}e), demonstrating strong correspondence on large and intermediate angular scales, with progressively reduced fidelity towards smaller scales. This decrease is expected despite the learned prior: small angular scales are dominated by shape noise and therefore weakly constrained by the data, so the posterior mean averages over the many small-scale configurations permitted by the prior, rather than reproducing the particular small-scale structure of each ground-truth field.\\

\noindent
We further test whether the reconstruction introduces a redshift-dependent suppression of density contrast. Using the same 16 held-out synthetic fields, we group the lens planes into the five redshift intervals used for the tomographic visualization, together with a sixth interval containing the remaining high-redshift planes.  For each field, the ground-truth and posterior mean density slices are smoothed identically with a 1-arcmin FWHM Gaussian, and we measure the standard deviation of the matter overdensity within the COSMOS-Web footprint. Although posterior averaging reduces the absolute contrast relative to the ground truth, the recovered fraction, defined as the ratio of the posterior-mean to ground-truth standard deviation, shows no systematic trend with redshift (Extended Data Fig.~\ref{fig:sdensity_contrast}). The reconstruction therefore does not introduce a strong additional redshift-dependent suppression of density contrast.\\

\noindent
Having tested the angular reconstruction, we separately assess recovery along the line of sight. We divide both the ground-truth and reconstructed 3D matter fields into six fixed tomographic slices and measure how structure originating in each ground-truth slice is distributed among the reconstructed slices (Extended Data Fig.~\ref{fig:synthetic_radial_recovery}). The first five slices span the same redshift intervals used for the five tomographic views of the COSMOS-Web reconstruction, while the sixth contains the five remaining high-redshift lens planes (Fig.~\ref{fig:3d_maps_slices}).
Let $X^{\rm truth}_{pa}(\boldsymbol\theta)$ and $X^{\rm rec}_{pb}(\boldsymbol\theta)$ denote the binned ground-truth and posterior mean fluctuations in validation field $p$, after subtracting the spatial mean of each binned map inside the common angular mask $\mathcal{M}$. For ground-truth slice $a$ and reconstructed slice $b$, we define the response
\begin{equation}
R^{(p)}_{ab} =
\frac{\sum_{\boldsymbol\theta\in\mathcal{M}}
X^{\rm truth}_{pa}(\boldsymbol\theta)
X^{\rm rec}_{pb}(\boldsymbol\theta)}
{\sum_{\boldsymbol\theta\in\mathcal{M}}
\left[X^{\rm truth}_{pa}(\boldsymbol\theta)\right]^2}.
\end{equation}
\noindent
We calculate the response separately in each validation field and report the median across the 16 fields (Extended Data Fig.~\ref{fig:synthetic_radial_recovery}a). A perfectly localized radial reconstruction would concentrate the response along the matrix diagonal, whereas radial mixing distributes the response across neighbouring reconstructed slices. The measured response is broad and extends into adjacent redshift intervals, as expected from the broad and overlapping WL efficiency kernels. This
test therefore directly quantifies the radial localization and line-of-sight
mixing of the reconstruction. \\

\noindent
\subsubsection*{Cosmological consistency}
We finally test whether the angular statistics of the COSMOS-Web reconstruction are consistent with those expected for the fiducial $\Lambda$CDM cosmology represented by \texttt{Conicus3D}. Each of the 1,024 COSMOS-Web posterior samples is projected into convergence using the source-weighted lensing kernel, and its angular band powers are measured with the flat-sky pseudo-$C_{\ell}$ estimator implemented in \texttt{NaMaster} \cite{Alonso2019}. We use five multipole bands with edges at $\ell$=500, 1,500, 2,500, 4,000, 6,000 and 8,000. To construct the cosmological reference distribution, we analyse 800 independent held-out \texttt{Conicus3D} fields from phase \texttt{ph000}, which is excluded from training. This ensemble is distinct from the 16 synthetic observations used above to test reconstruction fidelity: the 16-field analysis compares each reconstruction directly with its known ground truth, whereas the 800-field ensemble quantifies the field-to-field variation expected for COSMOS-Web-sized realizations of the fiducial cosmology. The same projection and measurement procedure is applied to all fields, which are analysed over the same C1-apodized COSMOS-Web footprint with identical treatment of mask-induced mode coupling, the pixel window and band-power binning. We characterize the posterior uncertainty and field-to-field variation of the simulated ensemble using the mean and central 68\% (16$^{\rm th}$--84$^{\rm th}$ percentile) intervals of their respective band-power distributions (Fig.~\ref{fig:recon_cosmology}).\\

\noindent
The fiducial theoretical convergence spectrum is computed for the cosmological parameters adopted by \texttt{Conicus3D} using the same source-weighted lensing kernel as for the reconstructed and simulated fields. We evaluate the non-linear matter power spectrum with \texttt{jax-cosmo} \cite{Campagne2023}, using the transfer function of \cite{Eisenstein1998} and the \texttt{halofit} prescription of \cite{Takahashi2012}, and project it onto the 20 lens planes in the Limber approximation. For direct comparison with the measured band powers, the theoretical spectrum is propagated through the same survey mask, mode-coupling and band-power operators. The resulting binned prediction therefore represents the fiducial $\Lambda$CDM spectrum as it would be measured over the COSMOS-Web footprint, rather than the corresponding ideal continuous spectrum. The analogous
fractional residual is defined relative to this binned fiducial prediction, rather than to a matched ground-truth realization as in the end-to-end validation above (Fig.~\ref{fig:recon_cosmology}b).\\

\noindent
To quantify the multivariate consistency of the reconstructed spectrum with the held-out simulation ensemble, we compare the five-dimensional vector of mean posterior band powers, $\mathbf{C}^{\rm post}$, with the mean band-power vector of the 800 \texttt{Conicus3D} realizations, $\overline{\mathbf{C}}^{\rm sim}$. We estimate the band-power covariance matrix, $\mathbf{\Sigma}$, from the same simulation ensemble and correct its inverse for finite-sample bias using the Hartlap factor \cite{Hartlap2007}. We then define the covariance-weighted squared distance
\begin{equation}
D^2 =
\left(
\mathbf{C}^{\rm post}
-
\overline{\mathbf{C}}^{\rm sim}
\right)^{\rm T}
\mathbf{\Sigma}_{\rm H}^{-1}
\left(
\mathbf{C}^{\rm post}
-
\overline{\mathbf{C}}^{\rm sim}
\right),
\end{equation}
where $\mathbf{\Sigma}_{\rm H}^{-1}$ denotes the Hartlap-corrected inverse covariance. For the COSMOS-Web reconstruction, we obtain $D^2=1.70$.\\

\noindent
We calibrate the significance of this statistic empirically using the held-out simulations rather than assuming a parametric distribution for $D^2$. To construct the empirical reference distribution, each \texttt{Conicus3D} realization is removed in turn, and its covariance-weighted distance is evaluated relative to the mean and covariance estimated from the remaining 799 realizations, with the corresponding Hartlap correction applied to the inverse covariance. The empirical $p$-value is defined as the fraction of these leave-one-out distances that are equal to or larger than the value measured for the COSMOS-Web reconstruction. This procedure yields $p=0.805$, indicating that the reconstructed band-power vector is statistically consistent with the distribution of COSMOS-Web-sized fields in the held-out simulations. Because the generative prior is itself learned from simulations of the same fiducial cosmology, this comparison is a consistency test of the reconstructed field rather than an independent cosmological constraint.\\

\noindent
We therefore also test the reconstruction when the true matter field is drawn from a cosmology other than that of the training simulations. We construct light cones in the same way as \texttt{Conicus3D} from AbacusSummit simulations \cite{hadzhiyska2022halo} in two alternative $\Lambda$CDM cosmologies with lower and higher clustering amplitude than the fiducial cosmology: \texttt{c001}, with $\Omega_{\rm m}=0.2761$, $h=0.7030$ and $\sigma_8=0.7768$, and \texttt{c015}, with $\Omega_{\rm m}=0.2950$, $h=0.6880$ and $\sigma_8=0.8350$. For each cosmology, we select the light cone with the same sky position and initial-condition phases as one of the 16 fiducial validation fields, generate a synthetic WL catalogue as described above and reconstruct it using the prior trained on the fiducial cosmology, drawing 1,024 posterior samples. Only the matter field differs from the fiducial validation; the lensing projection, source sampling, redshift uncertainties and shape noise are unchanged, so the test isolates the effect of a misspecified prior. We compare the posterior band powers with those of the ground-truth field and with the binned $\Lambda$CDM prediction of the fiducial cosmology used to train the prior, computed as described above (Extended Data Fig.~\ref{fig:synthetic_cosmology}). For both cosmologies, the mean posterior band powers lie within 16\% of the ground-truth values in every band, comparable to the fiducial validation (Extended Data Fig.~\ref{fig:PS_val}d). The \texttt{c015} ground-truth field has 42--97\% more power than this fiducial prediction (27--77\% more than the prediction of its own cosmology), and the mean posterior band powers remain 27--111\% above the fiducial prediction rather than reverting to it. These tests indicate that the recovered convergence power is set by the lensing data rather than by the cosmology of the training simulations.\\

\subsection*{Residual PSF Systematics}
\label{PSF_Validation}
Residual anisotropy in the point-spread function (PSF) can introduce coherent alignments into galaxy-shape measurements and thereby mimic WL structure. We test for residual PSF contamination at two complementary levels: coupling between the residual stellar PSF ellipticity
and the reconstructed shear field, and spatial and scale-dependent
correlations between PSF-induced and reconstructed convergence fields. The
stellar PSF residual catalogues for F115W and F150W are taken from the
COSMOS-Web WL analysis of \cite{Scognamiglio2026}, where the PSF was modelled for individual exposures and combined to account for spatial variations across the mosaic, including discontinuities at exposure boundaries. The residuals were then measured directly from stars across the survey footprint.\\

\noindent
We first test whether residual stellar PSF ellipticity is correlated with the
reconstructed shear field. The posterior-mean convergence reconstruction is converted into shear components, $\gamma^{\rm post}_1$ and
$\gamma^{\rm post}_2$, using the same angular grid as the reconstructed matter field. The F115W and F150W stellar PSF residual ellipticities are binned onto this grid. For the combined F115W+F150W test, we average the residual PSF ellipticity fields from the two bands, matching the equal-weight combination used for galaxies with shape measurements in both filters in the WL
catalogue \cite{Scognamiglio2026}. For each shear component, we fit
\begin{equation}
\gamma_i^{\rm post}
=
c_i+\alpha_i e_i^{\rm PSF,res},
\end{equation}
where $e_i^{\rm PSF,res}$ is the corresponding component of the residual
stellar PSF ellipticity, $c_i$ is a constant offset and $\alpha_i$ quantifies its linear coupling to the reconstructed shear. For the combined
F115W+F150W residual field, we obtain $\alpha_1=0.020$ and $\alpha_2=-0.047$. We assess the significance of these slopes using spatial shift-null realizations of the PSF residual field, which preserve its spatial structure while disrupting its alignment with the reconstructed shear. The measured slopes are consistent with the corresponding null distributions, with $p_{\rm shift}=0.73$ and $0.42$ for the two shear components, respectively (Extended Data Fig.~\ref{fig:PS_val_II}).\\

\noindent
We next test whether residual PSF structure is spatially correlated with the reconstructed convergence field. The stellar residual ellipticity fields are converted into PSF-induced convergence maps, $\kappa_{\rm PSF}$, and compared with the posterior convergence field, $\kappa_{\rm post}$. As above, the
combined F115W+F150W PSF field is formed by averaging the two filter-specific fields. We construct an empirical null distribution by translating $\kappa_{\rm PSF}$ by random spatial offsets relative to the fixed posterior field. These shifts disrupt the spatial correspondence between the two fields while preserving the morphology and autocorrelation of the PSF residual
pattern. For the combined F115W+F150W field, the zero-lag normalized
correlation is $r=-0.004$, consistent with the shift-null distribution
($p_{\rm shift}=0.992$). The rms amplitude of the PSF-induced convergence field is approximately $6\%$ of that of the posterior convergence field.\\

\noindent
Finally, we test for scale-dependent correlations between the reconstructed
and PSF-induced convergence fields using their angular cross-power spectrum and the normalized cross-correlation coefficient
\begin{equation}
r_\ell =
\frac{
C_\ell^{\kappa_{\rm post}\times\kappa_{\rm PSF}}
}{
\sqrt{
C_\ell^{\kappa_{\rm post}\times\kappa_{\rm post}}
C_\ell^{\kappa_{\rm PSF}\times\kappa_{\rm PSF}}
}
}.
\end{equation}
We compare these statistics with empirical confidence intervals obtained from 300 spatial shift-null realizations (Extended Data Fig.~\ref{fig:K_psf_cross_cor}). Across the angular scales tested, the measured cross-correlation remains consistent with the shift-null
distribution, with no statistically significant scale-dependent correlation
between the reconstructed convergence field and the residual PSF pattern.\\

\noindent
Together, these tests find no statistically significant evidence of the
observational PSF residuals measured in the COSMOS-Web imaging accounting for
the coherent structure recovered in the WL reconstruction.

\section*{Data Availability}
The JWST data (Programme ID: General Observer 1727) are publicly available at \url{https://exchg.calet.org/cosmosweb-public/DR0.5/}.\\
The HST data (Programme IDs: General Observer 9822 and 10092) are publicly available at \url{https://irsa.ipac.caltech.edu/data/COSMOS/}.\\
The XMM-\textit{Newton} dataset (Programme ID: 020336) is publicly available in
staged releases through the IPAC/IRSA website \url{https://irsa.ipac.caltech.edu/data/COSMOS/}.\\
The \textit{Chandra} data (Programme ID: 901037) are publicly available through the Chandra Data Archive at \url{https://cda.harvard.edu/chaser/}.\\

%\section{Code availability}
%The codes for xx

\section*{Acknowledgments}
D.S. carried out this research both at Duke University and at the Jet Propulsion Laboratory, California Institute of Technology. Part of this work was done by D.S., O.D., K.B., B.Z. at Jet Propulsion Laboratory, California Institute of Technology, under a contract with the National Aeronautics and Space Administration (80NM0018D0004) and funded through the President’s and Director’s Research and Development Fund (PDRDF). This work was supported in part by the OpenUniverse effort, which is
funded by NASA under JPL Contract Task 70-711320,``Maximizing
Science Exploitation of Simulated Cosmological Survey Data Across
Surveys". B.Z. and K.L.B. additionally acknowledge support from the Stanback Innovation Fund, an Amazon AI4Science Discovery Award, a Sloan Research Fellowship, and generous funding from Jim and Elaine Peterson. Support for this work was also provided by NASA grants JWST-GO-01727 and HST-AR15802 awarded by the Space Telescope Science Institute, operated by the Association of Universities for Research in Astronomy, Inc., under NASA contract NAS 5-26555. G.L. and R.M. acknowledge support from STFC via grant ST/X001075/1, the UK Space Agency via grant ST/W002612/1, and Innovate UK (grant no. TS/Y014693/1). D.H. was supported by the Swiss State Secretariat for Education, Research and Innovation (SERI) under contract number 521107294. This project has received funding from the European Union’s Horizon 2020 research and innovation programme under the Marie Skłodowska-Curie grant agreement No. 101148925. French COSMOS team members are partly supported by the Centre National d’Études Spatiales (CNES). O.I. acknowledges funding from the French Agence Nationale de la Recherche for the project iMAGE (grant ANR-22-CE31-0007).

\section*{Author Contributions Statement}
B.Z. and K.B. developed the three-dimensional reconstruction method, with B.Z. leading its implementation and application to JWST data. D.S. designed and led the validation and analysis procedures used to assess the quality of the three-dimensional reconstructions. D.S., B.Z., O.D., and K.B. contributed to the analysis and interpretation of the results. D.S. led the writing of the manuscript, with contributions from B.Z. and input and feedback from all authors. D.S., D.H., and G.L. produced the weak lensing shear catalogue and constructed the two-dimensional mass map. G.G and G.T. provided the optical spectroscopic redshifts and the galaxy cluster catalog, while M.S., O.I. produced the photometric catalogues used in the analysis.
% D.S. led and coordinated the project and, together with D.H. and G.L., produced the weak lensing shear catalogue and constructed the two-dimensional mass map. B.Z. developed the three-dimensional reconstruction pipeline under the supervision of K.B.. D.S. and B.Z. performed the analysis of the three-dimensional reconstruction, and D.S., B.Z., O.D. and K.B. contributed to the interpretation of the results. D.S. led the writing of the manuscript, with contributions from B.Z., and with input and comments from all authors. G.T. provided the optical spectroscopic redshifts, and M.S., O.I., H.B.A., J.R.W. and L.P. produced the photometric catalogues used in this analysis.

\subsection*{Corresponding author}
Correspondence to \href{mailto:dianas@jpl.nasa.gov}{Diana Scognamiglio}.

%\section*{Ethics declaration}
% The COSMOS team has a strong history of supporting diversity and fostering an inclusive workplace to ensure all current and prospective members of the organizations are valued for the unique contributions they bring to support program requirements at their highest levels of performance, and are committed to the vision represented by NASA's core value of inclusion in project activities throughout the full life cycle. At the institutional level, Diversity and Inclusion efforts fall into three main categories: bring practices, community outreach, and training. All the scientific institution involved in this project advance their goals through regular training, issuance of guidelines to all partner institutions, regular collection and reporting of metrics, issuance of anonymous surveys, and the establishment of well-defined lines of authority to align practices with diversity and inclusion objectives.

\section*{Competing Interests Statement}
The authors declare no competing interests.\\

\clearpage
% The bibliography is embedded below in the submitted manuscript.
% \bibliography{sn-bibliography}

\onecolumn
\section*{Extended Data} \label{sec:ext_figs}
\setcounter{figure}{0}
\renewcommand{\figurename}{Extended Data Fig.}

%%%Projected lightcone validation %%%
\noindent
\begin{minipage}{\textwidth}
    \centering
    \includegraphics[width=\textwidth]{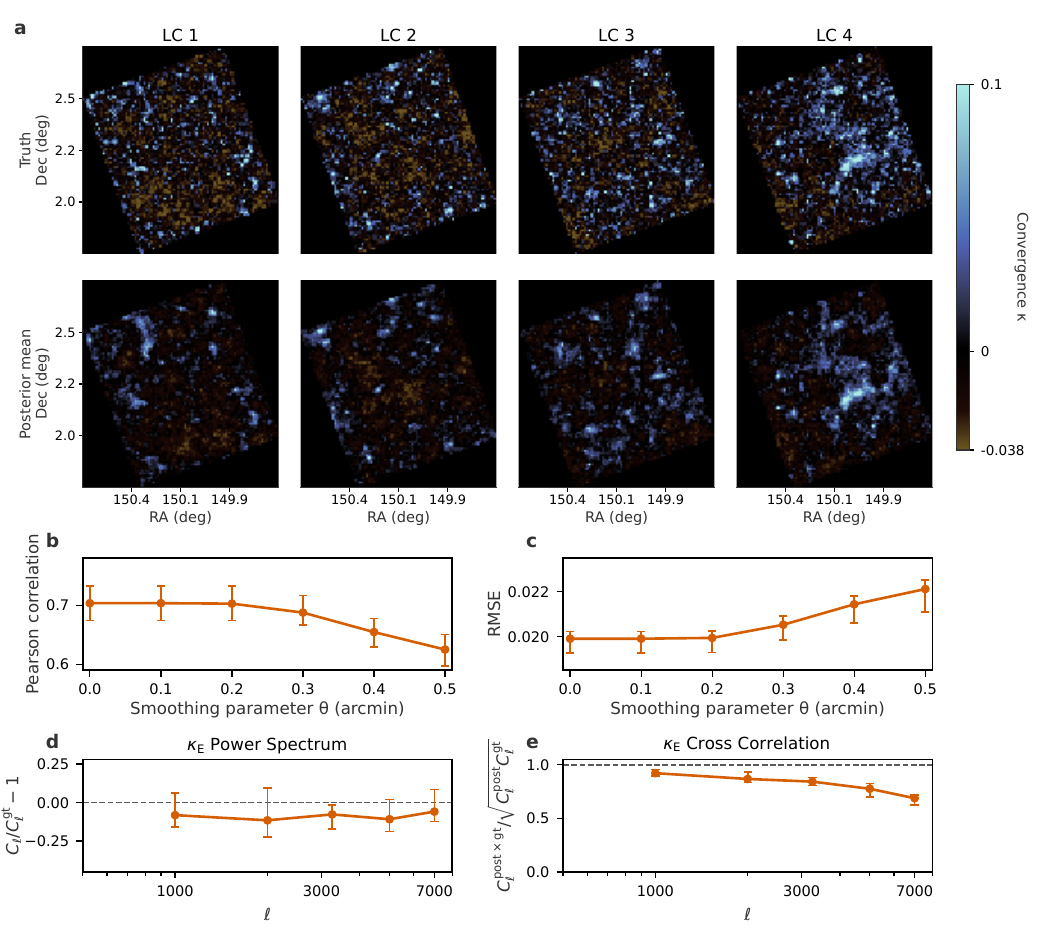}
    \captionof{figure}{\textbf{End-to-end reconstruction fidelity on held-out simulated fields.} \textbf{a.} Ground-truth and posterior mean projected convergence maps for four randomly selected examples from the 16 held-out fields. \textbf{b., c.} Pearson correlation and root-mean-square error (RMSE), respectively, across the 16 light cones as functions of the Gaussian standard deviation $\theta$ applied only to the posterior mean maps; the ground-truth maps remain at the native 0.6375-arcmin resolution. At native resolution, the median correlation is 0.70 and the median RMSE is 0.0199; neither metric improves with additional smoothing. Orange lines show field medians and error bars show 95\% paired field-bootstrap intervals. \textbf{d.} Fractional band-power residual, defined as the mean posterior band power divided by the matched ground-truth band power for each held-out field, minus unity. The median residual ranges from $-0.12$ to $-0.06$ across the five angular bands, indicating modest suppression of the recovered power amplitude. \textbf{e.} Normalized Fourier cross-power between each ground-truth map and its posterior mean. The median correlation decreases from 0.92 at $\ell=1{,}000$ to 0.69 at $\ell=7{,}000$, consistent with the higher lensing signal-to-noise ratio at lower multipoles. In \textbf{d, e}, orange lines show field medians and brackets show the central 68\% field ranges; dashed lines mark zero and unity in panel d. and e., respectively. }
    \label{fig:synthetic_projected_recovery}
    \label{fig:synthetic_scale_calibration}
    \label{fig:PS_val}
\end{minipage}
\clearpage

%%%%%Density contrast%%%%
\begin{figure*}[p]
    \centering
    \includegraphics[width=\textwidth]{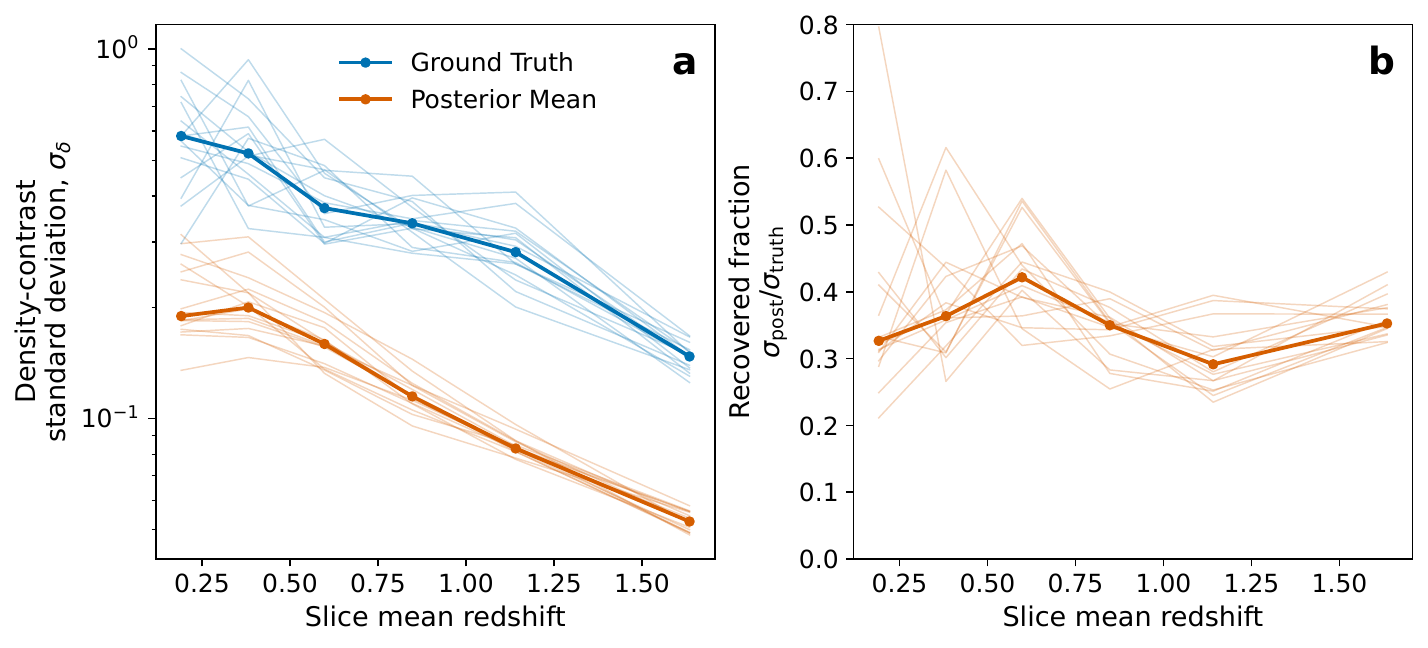}
    \caption{\textbf{
    Redshift dependence of density-contrast recovery in held-out simulations. a.} Standard deviation of the matter overdensity within the COSMOS-Web footprint as a function of mean slice redshift for 16 held-out synthetic fields. Lens planes are grouped into the five redshift intervals used for the tomographic reconstruction and a sixth higher-redshift interval containing the remaining lens planes. Ground-truth (blue) and posterior-mean reconstructed (orange) density fields are smoothed identically with a 1-arcmin FWHM Gaussian. Thin lines show individual fields and thick lines show the median across the 16 fields. \textbf{b.} Ratio of the posterior-mean to ground-truth standard deviation for the same fields. Posterior averaging suppresses the absolute density contrast relative to the ground truth, while the recovered fraction shows no evident systematic trend with redshift over the range tested.}
    \label{fig:sdensity_contrast}
\end{figure*}

%%% Radial validation %%%
\begin{figure*}[p]
    \centering
    \includegraphics[width=\textwidth]{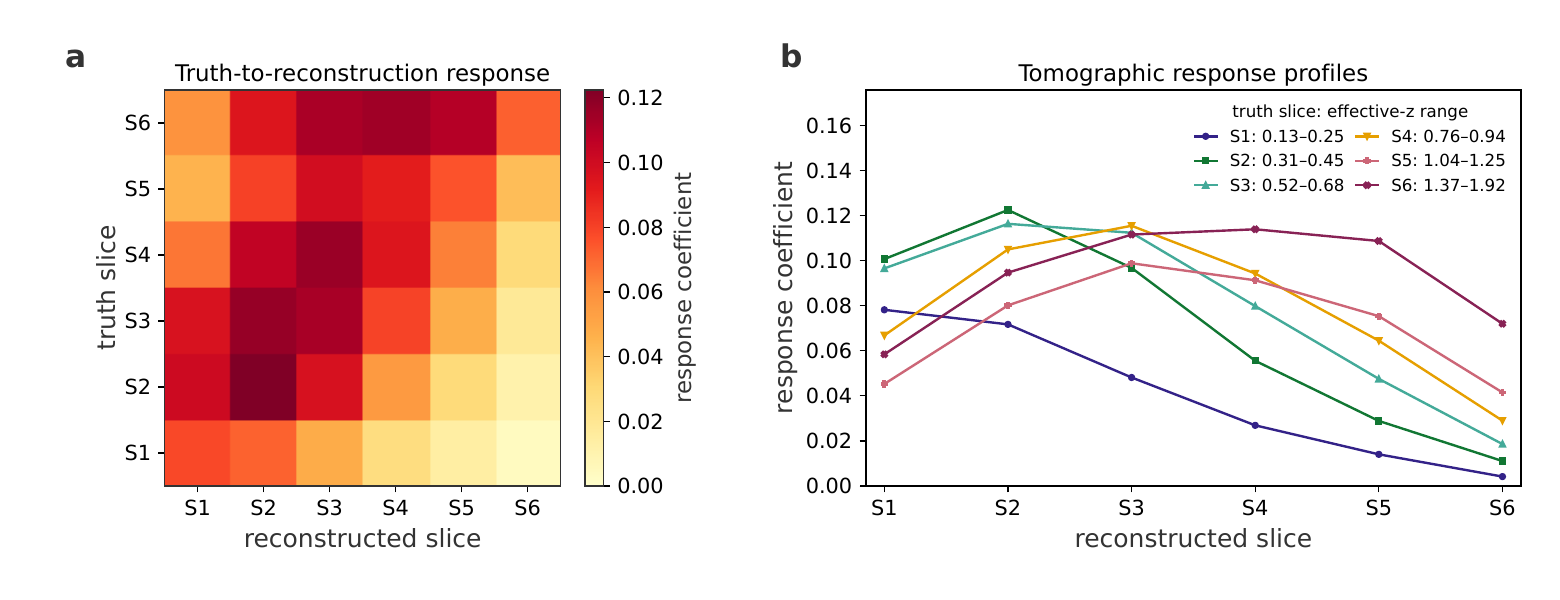}
    \caption{\textbf{
    %Truth-normalized radial response across six fixed tomographic slices.
    Radial response of the three-dimensional reconstruction in held-out simulations.} We quantify radial response on synthetic data using the five redshift slices shown in Fig.~\ref{fig:3d_maps_slices} and a sixth containing the remaining high-redshift planes. \textbf{a.} Median response matrix across the 16 held-out light cones. Each element \(R_{ab}\) quantifies the response of reconstructed slice \(b\) to structure in ground-truth slice \(a\), defined as their spatial cross-covariance divided by the variance of the ground-truth slice after mean subtraction within the common angular mask. Diagonal elements trace recovery within the corresponding redshift interval, while off-diagonal elements quantify radial mixing. \textbf{b.} Rows of the same response matrix, showing the reconstructed response to each ground-truth slice. The broad response across neighbouring slices reflects the broad, overlapping weak lensing efficiency kernels. No radial smoothing is applied.}
    \label{fig:synthetic_radial_recovery}
\end{figure*}

%%% Misspecified-cosmology validation %%%
\begin{figure*}[p]
    \centering
    \includegraphics[width=\textwidth]{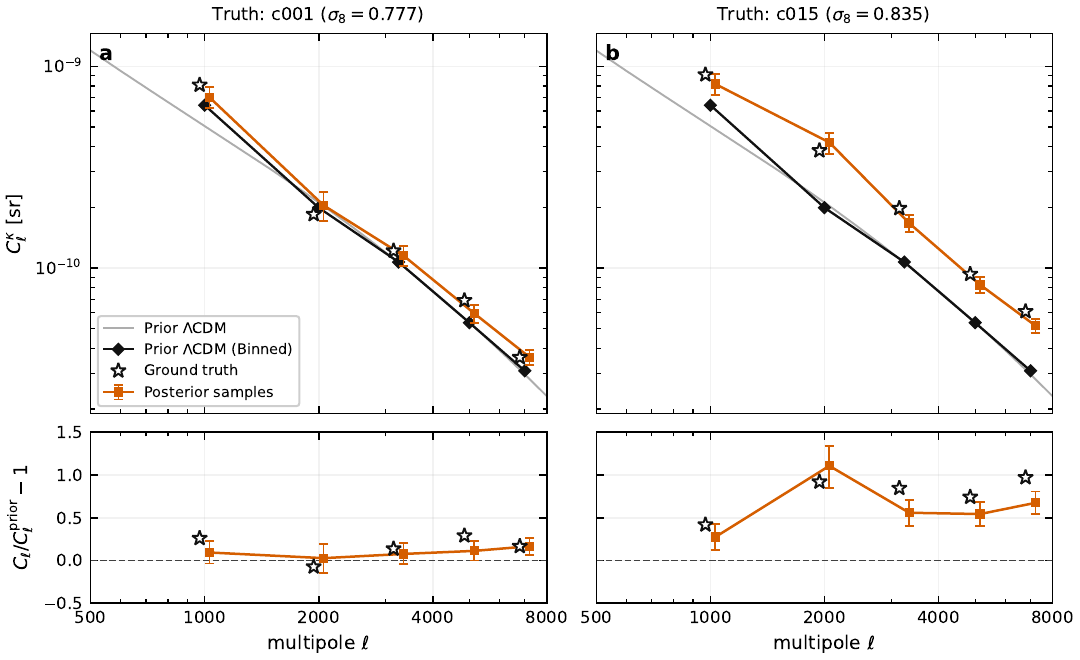}
    \caption{\textbf{Convergence power of held-out simulated fields from non-fiducial cosmologies. a., b.} Angular convergence band powers, $C_{\ell}^{\kappa}$, of a held-out field whose ground-truth matter distribution is drawn from AbacusSummit simulations in the \texttt{c001} ($\sigma_8=0.777$; \textbf{a.}) and \texttt{c015} ($\sigma_8=0.835$; \textbf{b.}) cosmologies and reconstructed with the generative prior trained on the fiducial cosmology ($\sigma_8=0.808$). Points and error bars show the mean and central 68\% (16th–84th percentile) interval across 1,024 posterior samples, and stars show the band powers of the ground-truth field. The continuous curve shows the $\Lambda$CDM spectrum of the fiducial cosmology used to train the prior; the binned prediction includes the same survey mask, mode-coupling and band-power operators applied to the reconstructed and ground-truth fields. Lower panels show fractional band-power residuals relative to this binned prior prediction. In both cosmologies, the reconstructed power follows that of the ground-truth field rather than the prior expectation, including the large excess power of the \texttt{c015} field.}
    \label{fig:synthetic_cosmology}
\end{figure*}

%%%PSF Validation II%%%%
\begin{figure*}[p]
    \centering  \includegraphics[width=\textwidth]{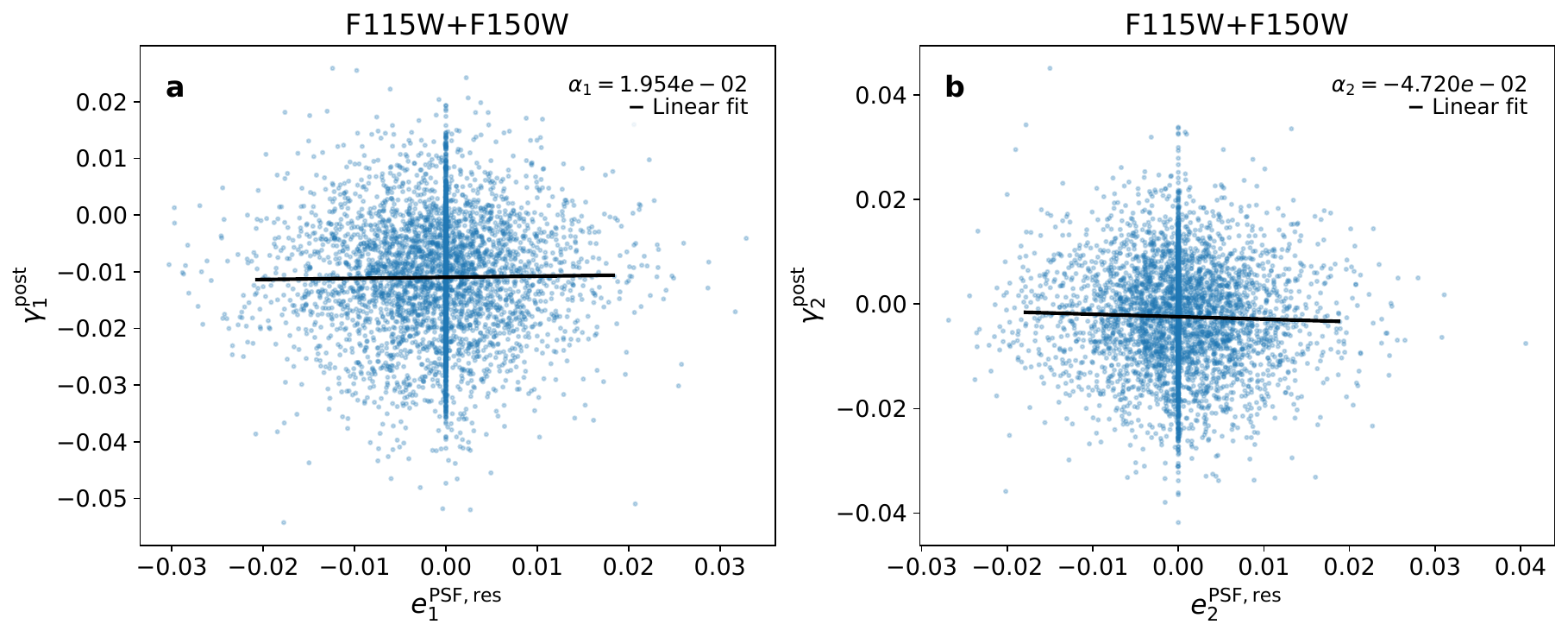}
\caption{\textbf{Test for residual PSF leakage into the reconstructed shear field. a., b.} Posterior mean shear components, $\gamma^{\rm post}_1$ and $\gamma^{\rm post}_2$, as a function of the corresponding residual stellar PSF ellipticity components, $e^{\rm PSF,res}_1$ and $e^{\rm PSF,res}_2$, for the combined F115W$+$F150W residual PSF field. The residual stellar ellipticities are binned onto the angular grid of the reconstructed shear field. Black lines show the best-fitting relations, $\gamma_i^{\rm post}=c_i+\alpha_i e_i^{\rm PSF,res}$. The measured leakage coefficients are $\alpha_1=0.020$ and $\alpha_2=-0.047$ and are consistent with the corresponding spatial shift-null distributions ($p_{\rm shift}=0.73$ and $0.42$, respectively). No statistically significant residual PSF leakage into the reconstructed shear field is detected.}  
    \label{fig:PS_val_II}
\end{figure*}
\clearpage
%%%PSF and WL maps CROSS CORRELATION %%%
\begin{figure*}[p]
 \includegraphics[width=\textwidth]{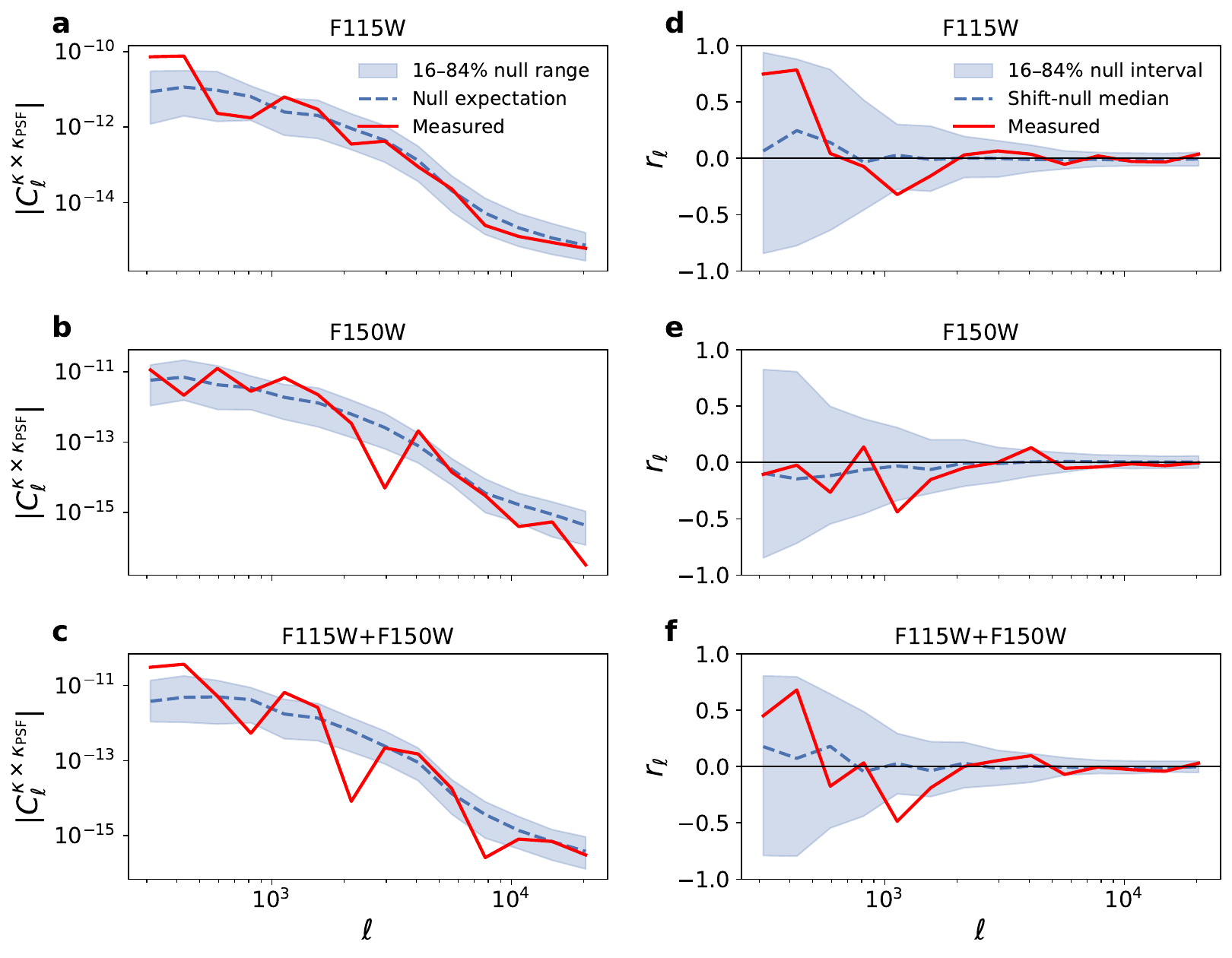}
    \caption{\textbf{Scale-dependent test for residual PSF contamination in the reconstructed convergence field. a–c} Absolute angular cross-power, $|C_\ell^{\kappa_{\rm post}\times\kappa_{\rm PSF}}|$, between the posterior convergence field, $\kappa_{\rm post}$, and PSF-induced convergence fields, $\kappa_{\rm PSF}$, constructed from the residual stellar PSF ellipticity fields in F115W, F150W and the combined F115W+F150W field, respectively. \textbf{d--f,} Corresponding normalized cross-correlation coefficients, $r_\ell$, retaining the sign of the cross-correlation. Red curves show the measured statistics, dashed blue curves show the median of 300 spatial shift-null realizations, and blue shaded regions show their 16th--84th percentile ranges. Across the multipole range tested, the measured correlations remain consistent with the shift-null distributions, with no statistically significant scale-dependent correlation between the reconstructed convergence field and residual PSF structure.
}
\label{fig:K_psf_cross_cor}
\end{figure*}

\end{document}